\documentclass[]{aa}
\usepackage{graphics}
\usepackage{graphicx}
\usepackage{txfonts}

\def\apj{ApJ}                 
\def\apjl{ApJ}                
\def\apjs{ApJS}               
\def\ao{Appl.~Opt.}           
\def\apss{Ap\&SS}             
\def\aap{A\&A}

\def\mnras{MNRAS}

\def\solphys{Sol.~Phys.}

\def\msol{M$_{\odot}$}
\def\lsol{L$_{\odot}$}
\def\ysol{Y$_{\odot}$}

\def\teff{T$_{eff}$}

\begin{document}      
%
   \title{Asteroseismology of solar-type stars: signatures of convective and/or helium cores.} 
%
 
        \titlerunning{Signatures of convective and/or helium cores.}  
 
   \author{
M.~Soriano
\and S.~Vauclair 
}
 
   \offprints{M. Soriano}

   \institute{Laboratoire d'Astrophysique de Toulouse-Tarbes - UMR 5572 - Universit\'e de Toulouse - CNRS, 14, av. E. Belin, 31400 Toulouse, France} 

\mail{sylvie.vauclair@ast.obs-mip.fr}
   
\date{Received \rule{2.0cm}{0.01cm} ; accepted \rule{2.0cm}{0.01cm} }

\authorrunning{ Soriano \& Vauclair}

\abstract
{}
{Several frequency combinations are widely used in the analysis of stellar oscillations for comparisons between models and observations. In particular, the ``small separations'' can help constraining the stellar cores. We showed in a previous paper that they can change sign, in contradiction with the ``asymptotic theory'', and that this behaviour could correspond to signatures of convective and/or helium cores. Here we analyse this behaviour in detail by systematic modelling during stellar evolution.}
{We computed evolutionary tracks for models with various masses (from 1.05 to 1.25 \msol) and various chemical compositions, with and without overshooting. We computed the adiabatic oscillation frequencies of the models and analysed the evolution of their small separations along an evolutionary track.}
{We find that, for all cases, the stars go through a stage during their evolution, where the small separations computed between degrees $\ell=0$~ and ~$\ell=2$ become negative in the observed range of frequencies. This behaviour is clearly related to the signature of a helium-rich core. We discuss the consequences for interpreting of the acoustic frequencies observed in solar-type stars.}
{}


\maketitle
                                                                                                                                         
\section{Introduction}

Asteroseismology has proved to be an excellent tool for deriving stellar parameters more precisely than ever before: mass, age, radius, metallicity, helium abundance value, stellar gravity, effective temperature. This has been discussed recently in detail for the case of the exoplanet-host star $\iota$~Hor (Vauclair et al. \cite{vauclair08}). Spectroscopic observations lead to values of \teff, $g$, and [Fe/H], which differ for different authors, according to the scale they use for the effective temperature determinations. The associated error bars are quite large. In this framework, observing the acoustic oscillations of the star and determining the mode frequencies is extremely helpful.

Once the oscillations of the star have been observed long enough (typically eight nights with HARPS), the analysis of the Fourier transform may lead to the determination of the ``large separations'', defined as $\Delta\nu_{n,\ell}=\nu_{n+1,\ell}-\nu_{n,\ell}$ and to the identification of the modes, as discussed in Bouchy et al. (\cite{bouchy05}). Meanwhile, various evolutionary tracks are computed with different input parameters (mass, chemical composition, presence or not of overshooting, etc). For a given set of parameters, only one model may reproduce the observed frequencies satisfactorily (Bazot et al. \cite{bazot05}; Soriano et al. \cite{soriano07}; Laymand \& Vauclair \cite{laymand07}; Vauclair et al. \cite{vauclair08}). As shown in Vauclair et al. (\cite{vauclair08}), the various models obtained in this way have similar gravities and ages. The other parameters are constrained with the help of the spectroscopic observational boxes. 

Asteroseismology can also give information about the internal structure of the stars and, more specifically, about the regions where the sound velocity changes rapidly. This happens in various transition layers, such as the limits of convective regions or layers with strong helium gradients. The transition layers that occur in stellar outer regions (bottom of outer convective zones, diffusion-induced helium gradients) may be characterised owing to the reflexion of the acoustic waves on the stellar surface and on the region of rapidly varying sound velocity. This is generally studied with the help of the ``second differences'', defined as $\delta_{2}\nu_{n,\ell}=\nu_{n+1,\ell}+\nu_{n-1,\ell}-2\nu_{n,\ell}$, which present oscillations with a period related to the acoustic depth of the transition layer (e.g. Gough \cite{gough90}; Monteiro \& Thompson \cite{monteiro98}; Vauclair \& Th\'eado \cite{vauclair04}; Castro \& Vauclair \cite{castro06}). On the other hand, the transition layers in the stellar internal regions are better characterised using the ``small separations'', defined as $\delta\nu_{n,\ell}=\nu_{n,\ell}-\nu_{n-1,\ell+2}$ (e.g. Roxburgh \& Vorontsov \cite{roxburgh94}). In a series of papers, Roxburgh \& Vorontsov (see Roxburgh \& Vorontsov \cite{roxburgh07} and references therein) show how the presence of a convective core could lead to oscillations in the small separations.

In a previous paper dedicated to the study of the exoplanet-host star HD 52265 (Soriano et al. \cite{soriano07}), we showed that in some cases the small separations, which should be positive in first approximation, could become negative. We explained how this special behaviour was related to either a convective core or a helium core with abrupt frontiers, resulting from the presence of a convective core in the past history of the star.

In the present paper, we give a general analysis of this effect for solar-type stars, with or without overshooting. We focused our analysis on stars that have either a solar metallicity or that are overmetallic, which is the case of exoplanet-host stars. We show that, along their evolutionary tracks, all stars go through a stage where the small separations between the $\ell= 0$ and $\ell= 2$ modes become negative, whether near the end of the main-sequence or at the beginning of the subgiant branch. This leads to a special behaviour in the ``echelle diagram'', which is obtained by plotting the mode frequencies in ordinates and the same frequencies modulo the large separations in abscissae (e.g. Soriano et al. \cite{soriano07}). Instead of remaining parallel, as predicted by the asymptotic theory (Tassoul \cite{tassoul80}), a crossing point appears between the $\ell=0$ and $\ell=2$ curves. This may also happen between the $\ell=1$ and $\ell=3$ curves, but it is less frequent and we concentrate here on the $\ell=0$ and $\ell=2$ case.

\begin{figure*}
\begin{center}
\includegraphics[angle=0,width=10cm]{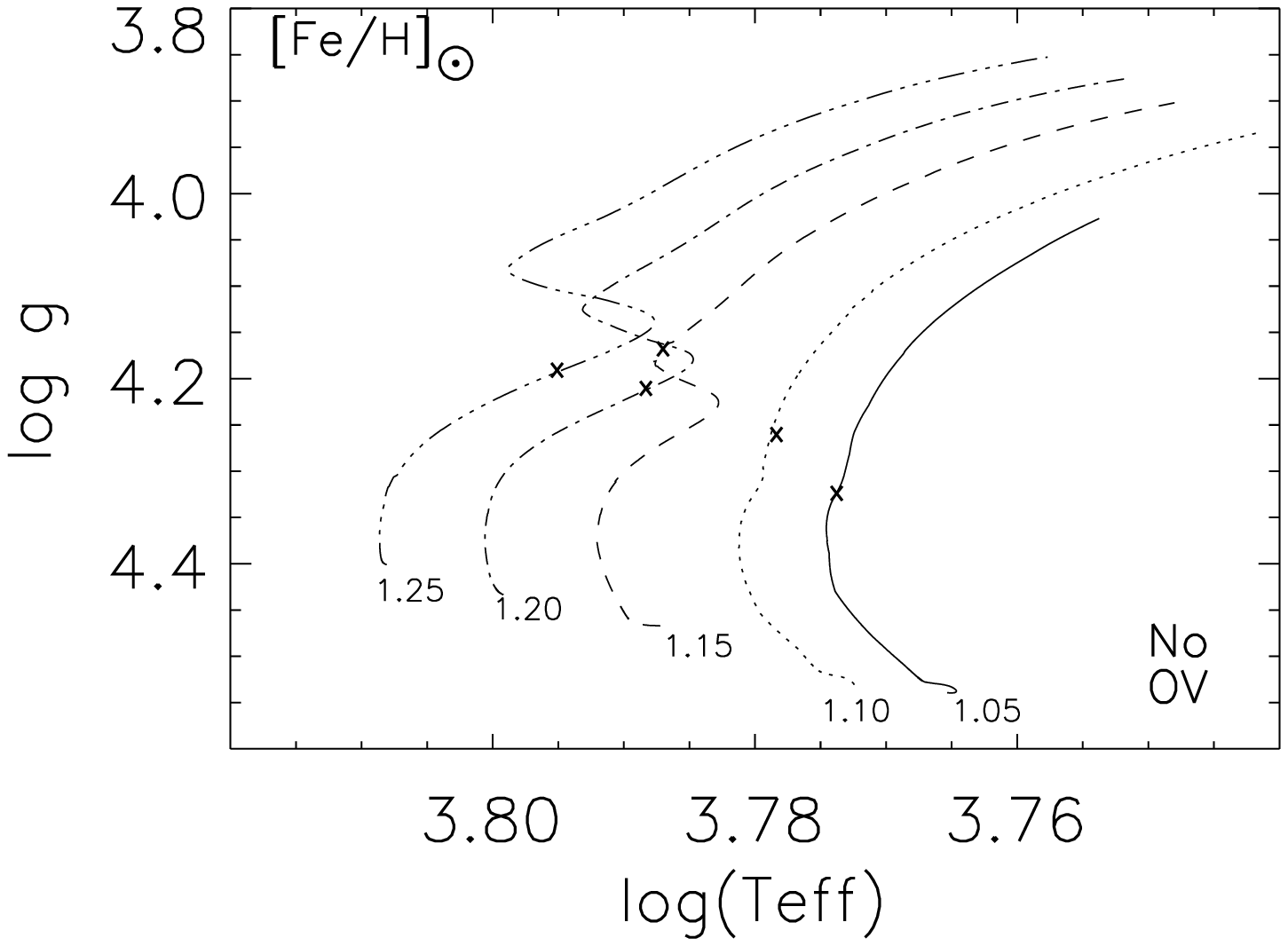}
\includegraphics[angle=0,width=10cm]{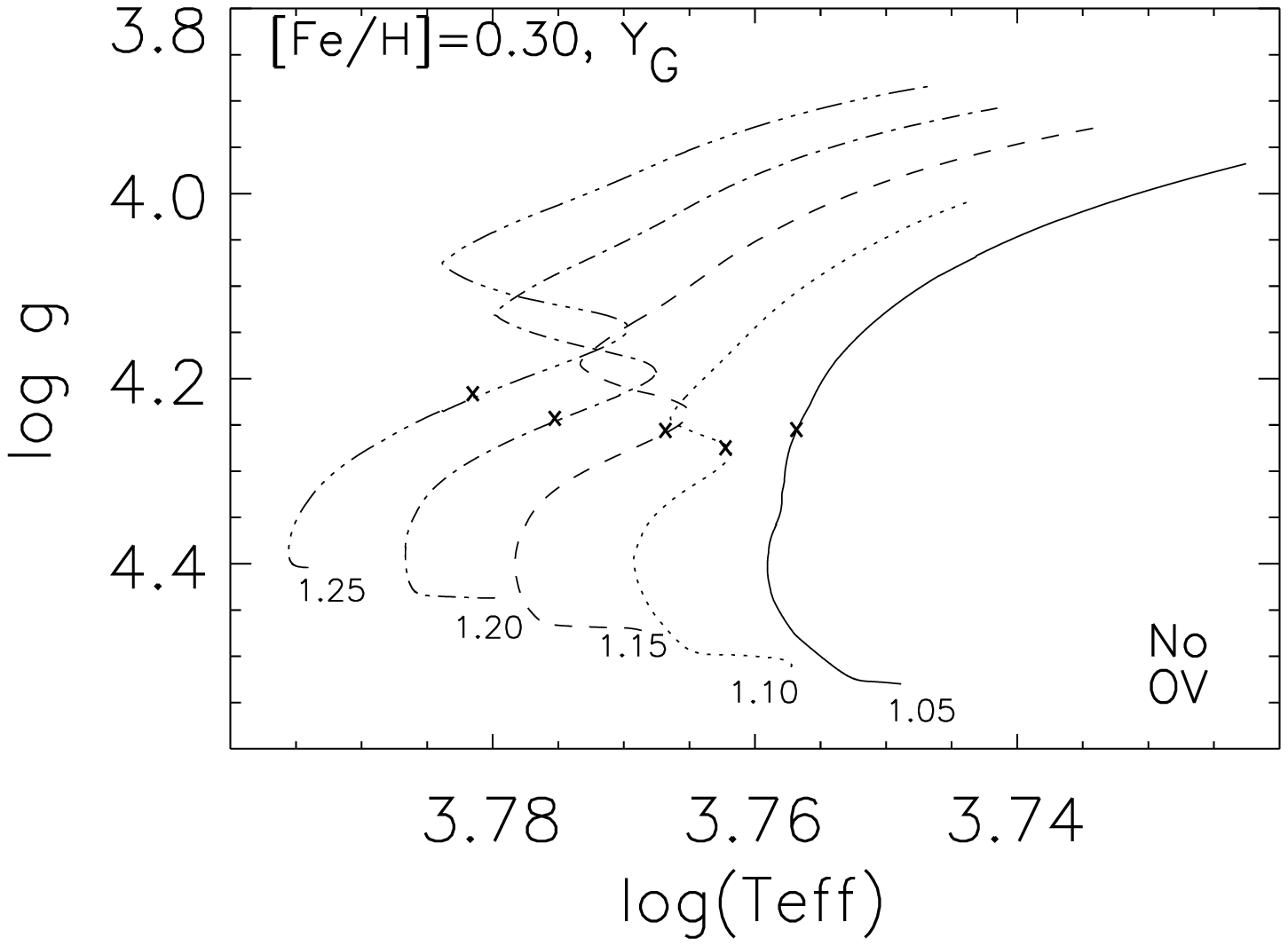}
\includegraphics[angle=0,width=10cm]{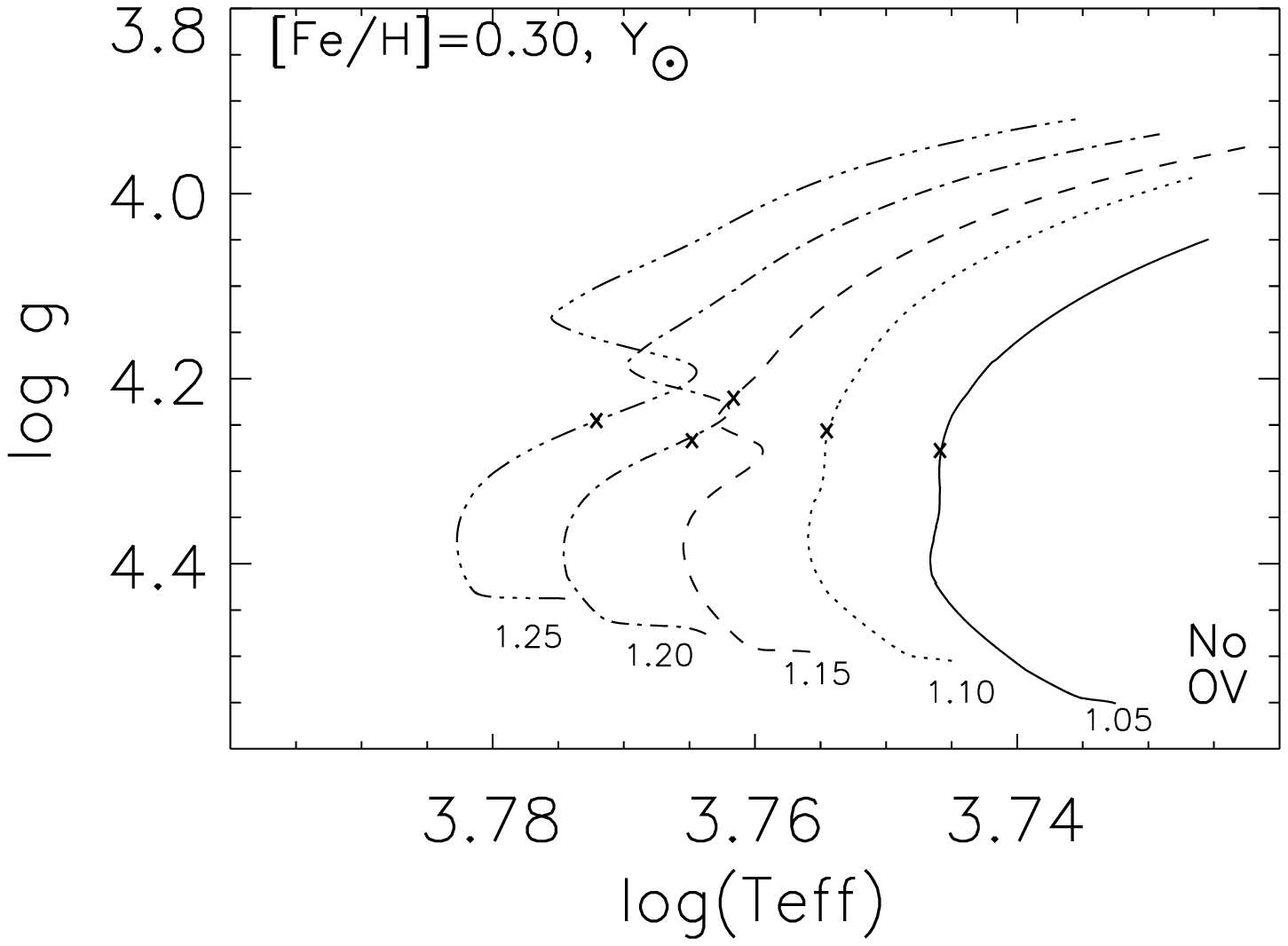}
\end{center}
\caption{Evolutionary tracks computed for solar metallicity [Fe/H]$_{\odot}$ (upper panel), overmetallicity [Fe/H]=0.30 with a helium abundance following the law of the chemical evolution of galaxies Y$_G$ (middle panel), and overmetallicity with a solar helium abundance \ysol (lower panel). All these tracks are computed without overshooting. The masses represented are: 1.05, 1.10, 1.15, 1.20, and 1.25 \msol.  The crosses represent the ``transition model'' for each evolutionary track, or first model from which we found negative small separations below 3.5 mHz. The characteristics of these models are displayed in Tables~\ref{tab1}, \ref{tab2}, and \ref{tab3}.}
\label{fig1}
\end{figure*}

\begin{figure*}
\begin{center}
\includegraphics[angle=0,width=10cm]{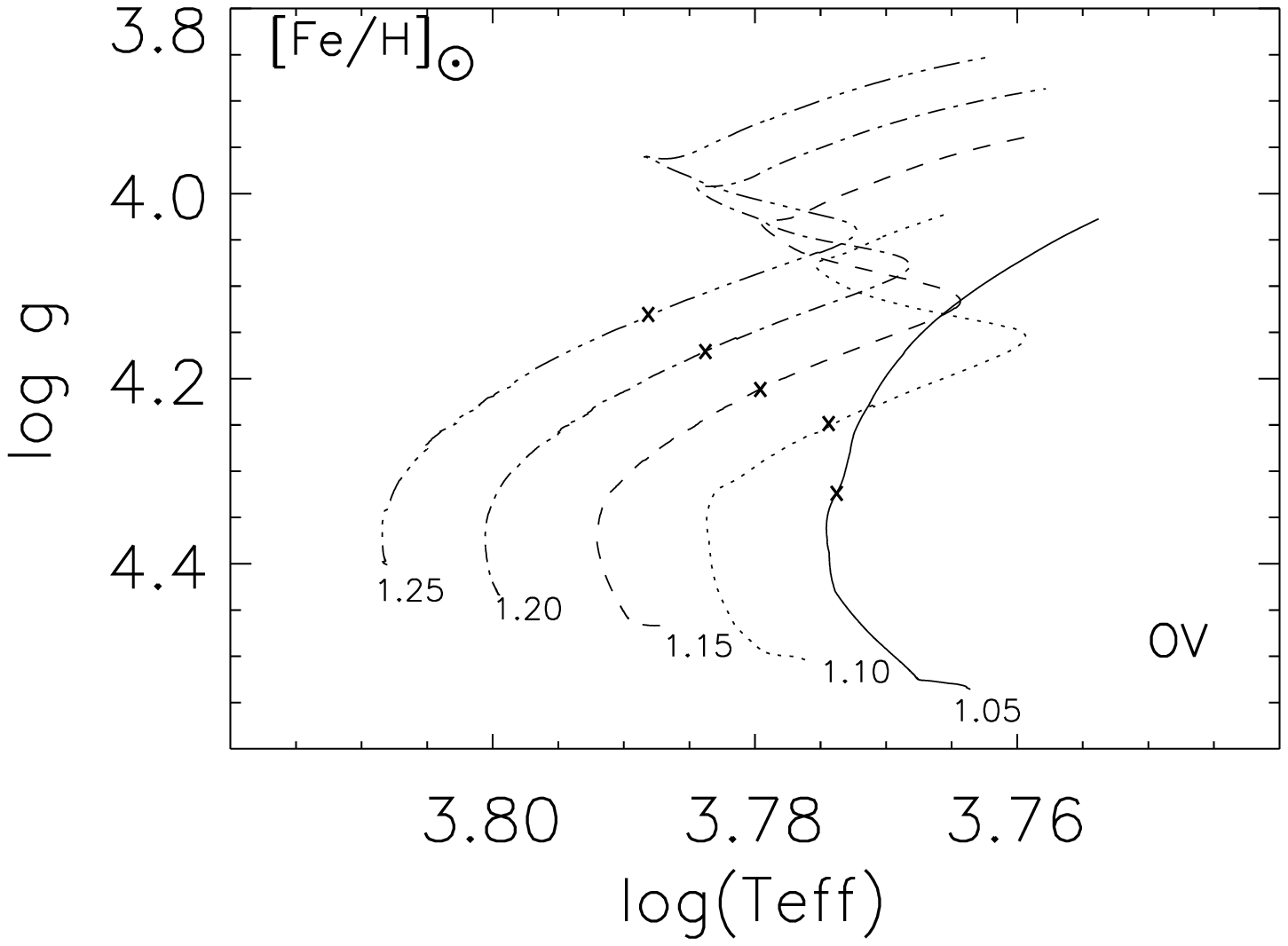}
\includegraphics[angle=0,width=10cm]{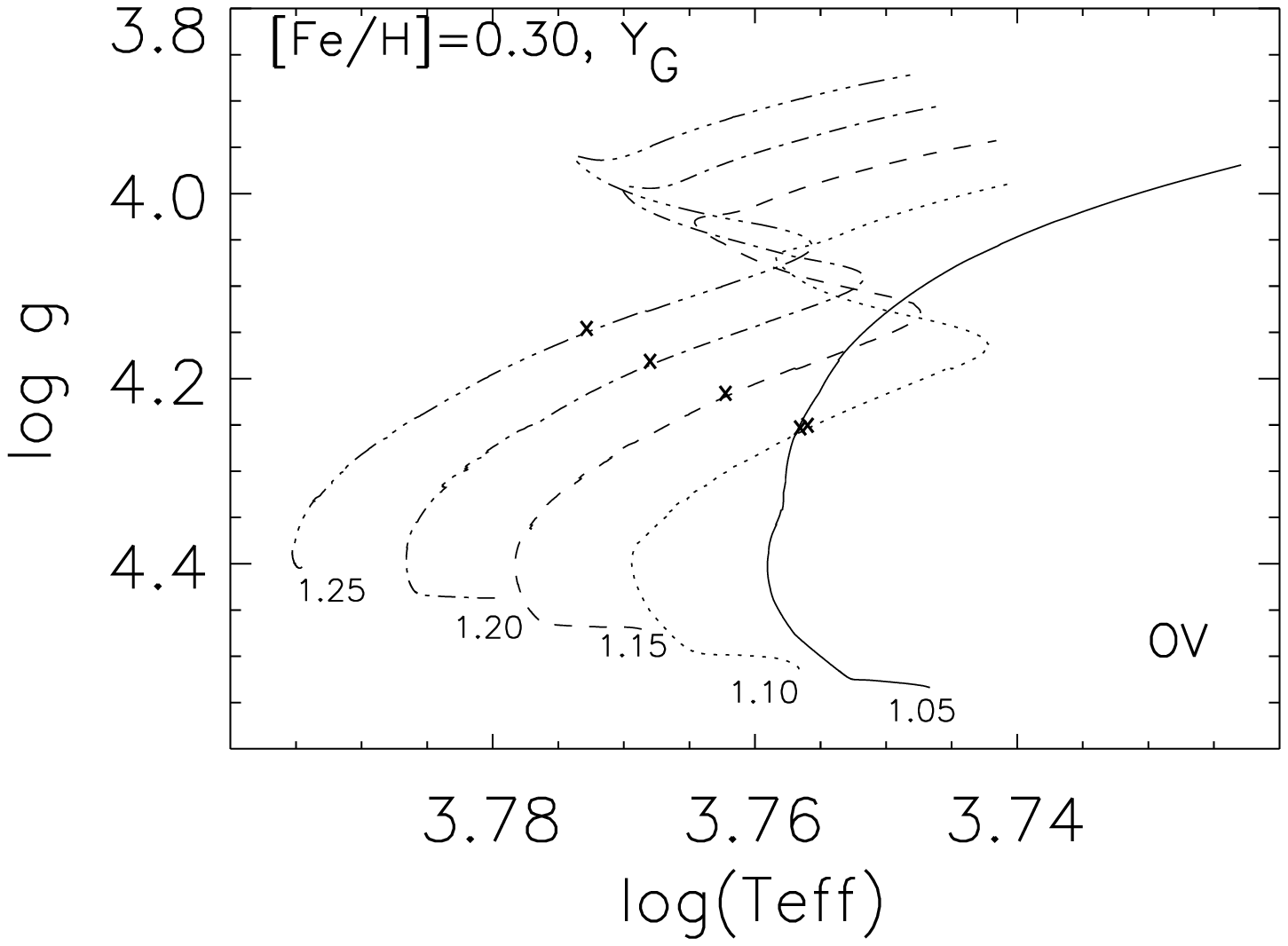}
\includegraphics[angle=0,width=10cm]{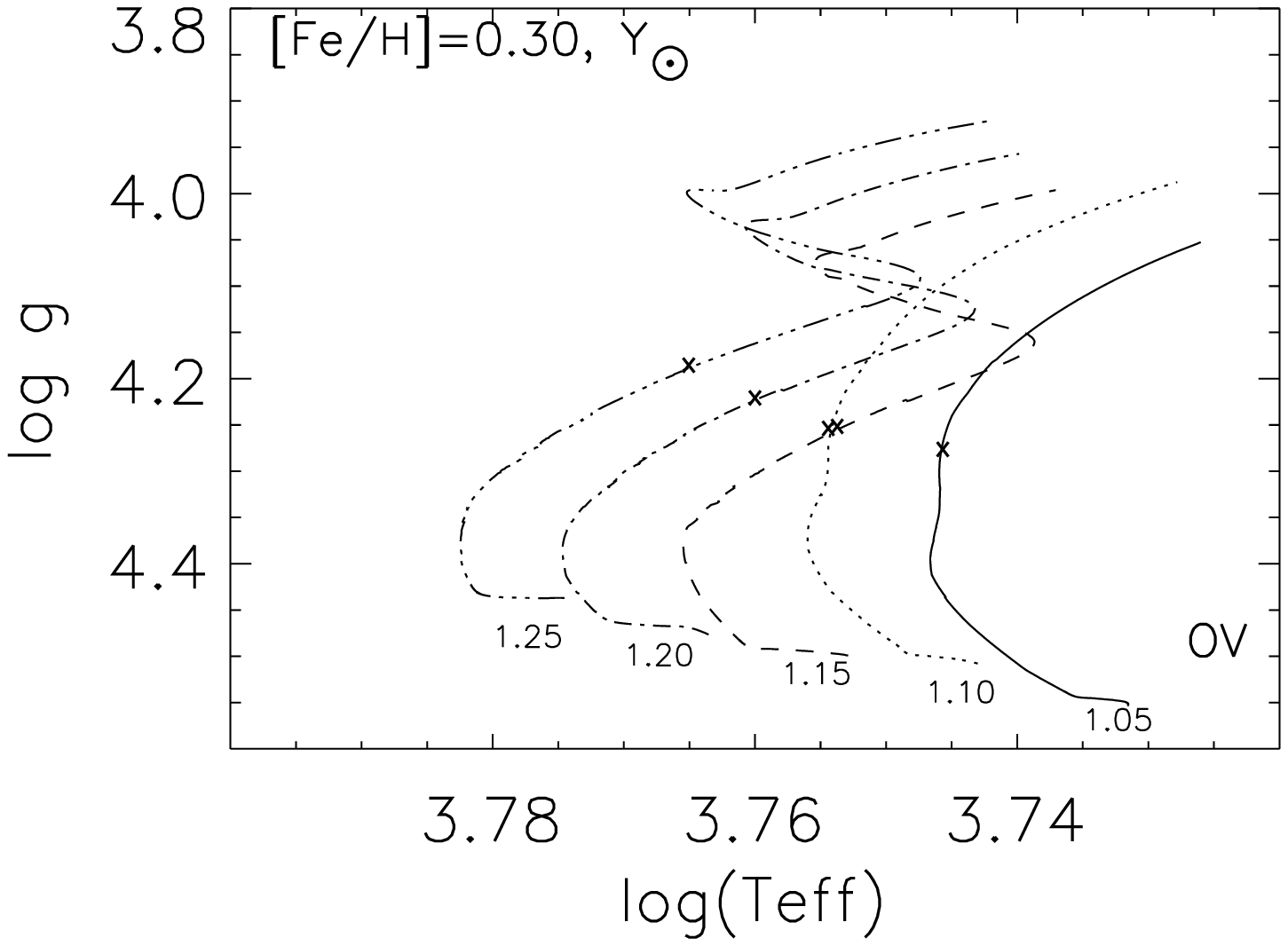}
\end{center}
\caption{Evolutionary tracks computed with the same hypothesis as in Fig.~\ref{fig1} but with overshooting. Here, the overshooting parameter is fixed as $\alpha_{ov}=0.20$. The masses represented are the same ones as in Fig.~\ref{fig1}. On each evolutionary track, we have represented the ``transition model'' by crosses, or first model from which we have negative small separations below 3.5 mHz. The characteristics of these models are displayed in Tables~\ref{tab4}, \ref{tab5}, and \ref{tab6}.}
\label{fig2}
\end{figure*}

For comparisons with observations, this behaviour becomes important if the crossing points correspond to frequencies below the cut-off frequency of the acoustic waves, which is the highest frequency for which the waves are reflected on the atmospheric regions. As this cut-off frequency depends on the characteristics of the stellar outer regions and may vary from star to star, we chose to study all the cases for which the crossing point occurs below 3.5 mHz.
Along the evolutionary track, we computed the models for which this behaviour appears for the first time, for various masses, and give their ages. In the following, we refer to these special models as ``transition models''. As the apparition of this behaviour is related to the radius of the convective or helium core, negative small separations appear earlier when overshooting is introduced. 

This particular behaviour of the small separations could be used to derive the size of stellar cores and the presence of overshooting for stars at the end of the main-sequence. A direct application to the exoplanet-host star $\mu$ Arae will be given in a forthcoming paper.

\section{Evolutionary tracks and models}

\begin{table*}
\caption{Parameters of the ``transition models'' computed without overshooting, with solar metallicity [Fe/H]$_{\odot}$ and solar helium value (Fig.~\ref{fig1}, upper panel).}
\label{tab1}
\begin{flushleft}
\begin{tabular}{cccccccccc} \hline
\hline
M$_{\star}$(\msol) &  Age (Gyr) & X$_C$       & Y$_C$  & R$_\star$(cm) & L$_\star$/\lsol & log \teff  & log $g$  & R$_{cc}$/R    & $t_{cc}$ (s) \cr
\hline \hline
1.05              &  7.3       & $<10^{-4}$  & 0.9758 & 9.07e10       & 1.854           & 3.771      & 4.332    & (0.07)$^{(a)}$& (124.4)$^{(b)}$ \cr
1.10              &  6.0       & $<10^{-4}$  & 0.9764 & 9.54e10       & 1.913           & 3.779      & 4.268    & (0.07)$^{(a)}$& (128.6)$^{(b)}$ \cr
1.15              &  4.9       & $<10^{-4}$  & 0.9785 & 10.12e10      & 2.681           & 3.788      & 4.176    & (0.06)$^{(a)}$& (127)$^{(b)}$   \cr
1.20              &  3.7       & 0.1462      & 0.8338 & 9.87e10       & 2.579           & 3.789      & 4.216    & 0.05         & 113             \cr
1.25              &  3.0       & 0.1986      & 0.7817 & 10.30e10      & 2.992           & 3.796      & 4.196    & 0.06         & 123.7           \cr
\hline
\end{tabular}
\end{flushleft}
$^{(a)}$ The values in parenthesis correspond to the cases where the convective core has disappeared, leaving a helium core with abrupt limits; here the values correspond to the linear radii of the helium cores.\\
$^{(b)}$ The values in parenthesis correspond to the acoustic radii of the helium cores.
\end{table*}

\begin{table*}
\caption{Same as Table~\ref{tab1} for the overmetallic models ([Fe/H]=0.30) with a helium mass fraction Y$_G$ (Fig.~\ref{fig1}, middle panel).}
\label{tab2}
\begin{flushleft}
\begin{tabular}{cccccccccc} \hline
\hline
M$_{\star}$(\msol) &  Age (Gyr) & X$_C$       & Y$_C$  & R$_\star$(cm) & L$_\star$/\lsol & log \teff  & log $g$  & R$_{cc}$/R     & $t_{cc}$ (s)   \cr
\hline \hline
1.05              &  7.6       & $<10^{-4}$  & 0.9567 & 8.75e10       & 1.511           & 3.757      & 4.263    & (0.07)        & (121.6)        \cr
1.10              &  5.7       & 0.0462      & 0.9173 & 8.79e10       & 1.594           & 3.761      & 4.279    & 0.05          & 98.2           \cr
1.15              &  4.7       & 0.1058      & 0.8588 & 9.15e10       & 1.819           & 3.767      & 4.263    & 0.06          & 113.1          \cr
1.20              &  3.7       & 0.1760      & 0.7893 & 9.50e10       & 2.117           & 3.776      & 4.250    & 0.06          & 126.4          \cr
1.25              &  3.1       & 0.2204      & 0.7451 & 9.98e10       & 2.481           & 3.782      & 4.224    & 0.07          & 140            \cr
\hline
\end{tabular}
\end{flushleft}
\end{table*}

\begin{table*}
\caption{Same as Table~\ref{tab1} for the overmetallic models ([Fe/H]=0.30) with a helium mass fraction \ysol (Fig.~\ref{fig1}, lower panel)}
\label{tab3}
\begin{flushleft}
\begin{tabular}{cccccccccc} \hline
\hline
M$_{\star}$(\msol) &  Age (Gyr) & X$_C$       & Y$_C$  & R$_\star$(cm) & L$_\star$/\lsol & log \teff  & log $g$ & R$_{cc}$/R      & $t_{cc}$ (s)   \cr
\hline \hline
1.05              &  9.7       & $<10^{-4}$  & 0.9545 & 8.52e10       & 1.294           & 3.746      & 4.285   & (0.06)         & (104.6)        \cr
1.10              &  7.9       & $<10^{-4}$  & 0.9565 & 8.94e10       & 1.536           & 3.754      & 4.264   & (0.06)         & (108.1)        \cr
1.15              &  6.6       & $<10^{-4}$  & 0.9609 & 9.52e10       & 1.877           & 3.762      & 4.229   & (0.06)         & (111.4)        \cr
1.20              &  4.9       & 0.1220      & 0.8427 & 9.30e10       & 1.848           & 3.765      & 4.267   & 0.05           & 108            \cr
1.25              &  4.0       & 0.1787      & 0.7864 & 9.70e10       & 2.146           & 3.773      & 4.249   & 0.06           & 122.5          \cr
\hline
\end{tabular}
\end{flushleft}
\end{table*}

\begin{table*}
\caption{Parameters of the ``transition models'' computed with overshooting, with solar metallicity [Fe/H]$_{\odot}$ and solar helium value (Fig.~\ref{fig2}, upper panel)}
\label{tab4}
\begin{flushleft}
\begin{tabular}{cccccccccc} \hline
\hline
M$_{\star}$(\msol) &  Age (Gyr) & X$_C$       & Y$_C$  & R$_\star$(cm) & L$_\star$/\lsol & log \teff  & log $g$ & R$_{cc}$/R     & $t_{cc}$ (s)   \cr
\hline \hline
1.05              &  7.3       & $<10^{-4}$  & 0.9758 & 9.07e10       & 1.854           & 3.771      & 4.332   & (0.7)$^{(a)}$ & (124.4)$^{(b)}$\cr
1.10              &  5.5       & 0.2285      & 0.7514 & 9.02e10       & 1.896           & 3.775      & 4.256   & 0.06         & 116.8    
      \cr
1.15              &  4.8       & 0.2369      & 0.7430 & 9.67e10       & 2.284           & 3.780      & 4.215   & 0.07         & 129.7          \cr
1.20              &  4.2       & 0.2429      & 0.7371 & 10.28e10      & 2.723           & 3.784      & 4.174   & 0.07         & 139.9          \cr
1.25              &  3.7       & 0.2472      & 0.7329 & 11.04e10      & 3.223           & 3.789      & 4.136   & 0.07         & 148.1          \cr
\hline
\end{tabular}
\end{flushleft}
$^{(a)}$ The values in parenthesis correspond to the cases where the convective core has disappeared, leaving a helium core with abrupt limits; here the values correspond to the linear radii of the helium cores.\\ 
$^{(b)}$ The values in parenthesis correspond to the acoustic radii of the helium cores.\end{table*}

\begin{table*}
\caption{ Same as Table~\ref{tab4} for the overmetallic models ([Fe/H]=0.30) with a helium mass fraction Y$_G$ (Fig.\ref{fig2}, middle panel).}
\label{tab5}
\begin{flushleft}
\begin{tabular}{cccccccccc} \hline
\hline
M$_{\star}$(\msol) &  Age (Gyr) & X$_C$       & Y$_C$  & R$_\star$(cm) & L$_\star$/\lsol & log \teff  & log $g$  & R$_{cc}$/R      & $t_{cc}$ (s)   \cr
\hline \hline
1.05              &  7.6       & $<10^{-4}$  & 0.9567 & 8.77e10       & 1.516           & 3.757      & 4.261    & (0.07)        & (121.6)        \cr
1.10              &  6.3       & 0.2107      & 0.7539 & 9.05e10       & 1.594           & 3.757      & 4.258    & 0.07          & 136.4          \cr
1.15              &  5.4       & 0.2179      & 0.7468 & 9.58e10       & 1.912           & 3.763      & 4.223    & 0.07          & 146.5          \cr
1.20              &  4.7       & 0.2232      & 0.7418 & 10.20e10      & 2.277           & 3.769      & 4.189    & 0.07          & 155.4          \cr
1.25              &  4.1       & 0.2293      & 0.7358 & 10.83e10      & 2.691           & 3.773      & 4.153    & 0.07          & 161.3          \cr
\hline
\end{tabular}
\end{flushleft}
\end{table*}

\begin{table*}
\caption{Same as Table~\ref{tab4} for the overmetallic models ([Fe/H]=0.30) with a helium mass fraction \ysol (Fig.~\ref{fig2}, lower panel).}
\label{tab6}
\begin{flushleft}
\begin{tabular}{cccccccccc} \hline
\hline
M$_{\star}$(\msol) &  Age (Gyr) & X$_C$       & Y$_C$  & R$_\star$(cm) & L$_\star$/\lsol & log \teff  & log $g$  & R$_{cc}$/R      & $t_{cc}$ (s)   \cr
\hline \hline
1.05              &  9.7       & $<10^{-4}$  & 0.9545 & 8.54e10       & 1.298           & 3.746      & 4.284    & (0.06)        & (104.4)        \cr
1.10              &  8.0       & $<10^{-4}$  & 0.9565 & 8.97e10       & 1.543           & 3.753      & 4.261    & (0.06)        & (107.7)        \cr
1.15              &  6.6       & 0.2267      & 0.7379 & 9.20e10       & 1.627           & 3.754      & 4.259    & 0.07          & 136.3          \cr
1.20              &  5.7       & 0.2347      & 0.7301 & 9.75e10       & 1.938           & 3.761      & 4.227    & 0.07          & 144.8          \cr
1.25              &  5.0       & 0.2402      & 0.7248 & 10.35e10      & 2.290           & 3.766      & 4.192    & 0.07          & 155.6          \cr
\hline
\end{tabular}
\end{flushleft}
\end{table*}

We computed series of evolutionary tracks using the Toulouse Geneva Evolution Code (see Hui Bon Hoa \cite{hui07} for a general description of this code) with the OPAL equation of state and opacities (Rogers \& Nayfonov \cite{rogers02}; Iglesias \& Rogers \cite{iglesias96}) and the NACRE nuclear reaction rates (Angulo et al. \cite{angulo99}).
In all our models, we included microscopic diffusion as described in Michaud et al. (\cite{michaud04}) and Paquette et al. (\cite{paquette86}).

The convection was treated in the framework of the mixing length theory and the mixing length parameter was adjusted as in the Sun: $\alpha=1.8$ (Richard et al. \cite{richard04}).
Models were computed for a range of masses from 1.05 to 1.25 \msol.

We computed six series of models. The first three series were computed without overshooting and the second three series with overshooting at the limit of the stellar core. Here overshooting is described as an extension of the central convective zone by a length $\alpha_{ov}H_P$, where $H_P$ is the pressure height scale, and $\alpha_{ov}$ the overshooting parameter. In our computations, we fixed $\alpha_{ov}$ to 0.20. In each case, the three series differ from their abundances:
\begin{itemize}
\item Metallic and helium solar values. For the metallic values, the ``old'' Grevesse \& Noels (\cite{grevesse93}) abundances, compatible with helioseismology, are used: $X_{ini}=0.7097$, $Y_{ini}=0.2714$, $Z_{ini}=0.0189$~(Fig.~\ref{fig1}, upper panel, and Table~\ref{tab1} for the case without overshooting; Fig.~\ref{fig2}, upper panel, and Table~\ref{tab4} for the case with overshooting).
\item Overmetallic abundance [Fe/H]=0.30, and helium value computed as for the chemical evolution of galaxies, namely $dY/dZ=2.8 \pm 0.5$: $X_{ini}=0.6648$, $Y_{ini}=0.3027$, $Z_{ini}=0.0324$ (Izotov \& Thuan \cite{izotov04}) (Fig.~\ref{fig1}, middle panel, and Table~\ref{tab2} for the case without overshooting; Fig.~\ref{fig2}, middle panel, and Table~\ref{tab5} for the case with overshooting).
\item Overmetallic abundance [Fe/H]=0.30, and the solar helium value \ysol: $X_{ini}=0.6961$, $Y_{ini}=0.2714$, $Z_{ini}=0.0325$~(Fig.~\ref{fig1}, lower panel, and Table~\ref{tab3} for the case without overshooting; Fig.~\ref{fig2}, lower panel, and Table~\ref{tab6} for the case with overshooting).
\end{itemize}

We computed adiabatic oscillation frequencies for a large number of models along each evolutionary track, using the PULSE code (Brassard et al. \cite{brassard92}). The oscillation frequencies were computed for degrees $\ell=0$ to $\ell=3$, which are the only degrees observable for solar-like stars, because of the lack of spatial resolution. We only kept frequencies between 1.5 and 3.5 mHz, corresponding to the typical observational range for solar-type stars. Their radial orders range between 4 to 100. Using these frequencies, we computed the small separations for each model and analysed in what conditions these quantities can become negative. For all evolutionary tracks, we found models where the small separations become negative at some frequencies, which means that the $\ell$ = 0 and $\ell$ = 2 lines cross over at a given point in the echelle diagram (cf Fig.~\ref{fig3}, for example). In all cases, the frequencies of these crossing points decrease for increasing stellar age. For each computed track, we picked up the ``transition model'' for which the frequency of the crossing point is 3.5 mHz. These models are represented by crosses in Figs.~\ref{fig1} and \ref{fig2}, and their characteristics are given in Tables~\ref{tab1} to \ref{tab6}.

\section{Analysis of the results}

A first analysis of the possibilities of negative small separations in stars was given in Soriano et al.(\cite{soriano07}) for the specific case of the exoplanet-host star HD~52265. When we computed models for this star, taking spectroscopic constraints into account, we were surprised to find two models that showed this specific behaviour: one model at the end of the main sequence with a mass of 1.31 \msol, which had a convective core, and one model at the beginning of the subgiant branch with a mass of 1.20 \msol, in which the convective core, present during the main sequence, had disappeared. For both cases we realised that the main reason for the negative small separations was related to the high helium content of the cores. The basic role of convection in this respect was its action of concentrating helium inside a sharp core during the main sequence.\\ 
We showed how negative small separations could be a signature of the size of a helium core, as well as that of a convective core when it is still present. 

We give below a more complete analysis of the results we obtained by doing systematic studies of this effect for solar-type stars. We first recall useful theoretical points, then we discuss our computational results.

\subsection{Theoretical analysis}

In all our models, the acoustic frequencies were computed precisely using the PULSE adiabatic code (Brassard et al. \cite{brassard92}). The results take the particular features of the stellar interiors into account. That the small separations become negative in some cases is real.

This seems surprising at first sight because it contradicts what is usually called ``asymptotic theory'', as developed by Tassoul (\cite{tassoul80}). According to this analytical description of the oscillations, the large separations should be constant, equal to half of the inverse of the acoustic time, defined as the time needed for the acoustic waves to cross the stellar radius. Meanwhile, the small separation should vary quite slowly and remain positive. Although these approximate expressions are not used in real computations, they are very useful for understanding the underlying physics.

\begin{figure*}
\begin{center}
\includegraphics[angle=0,totalheight=5.5cm,width=8cm]{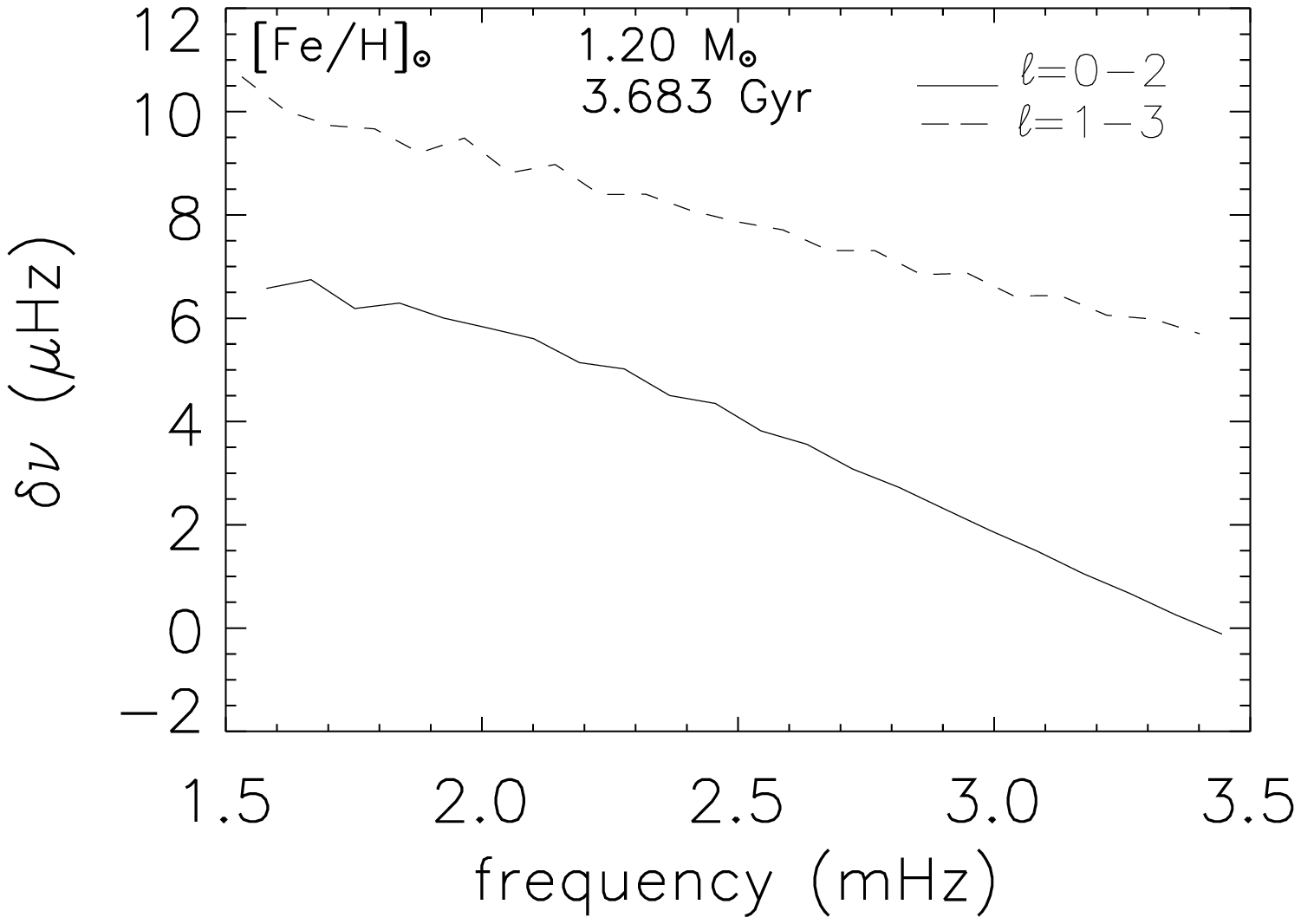}\includegraphics[angle=0,totalheight=5.5cm,width=8cm]{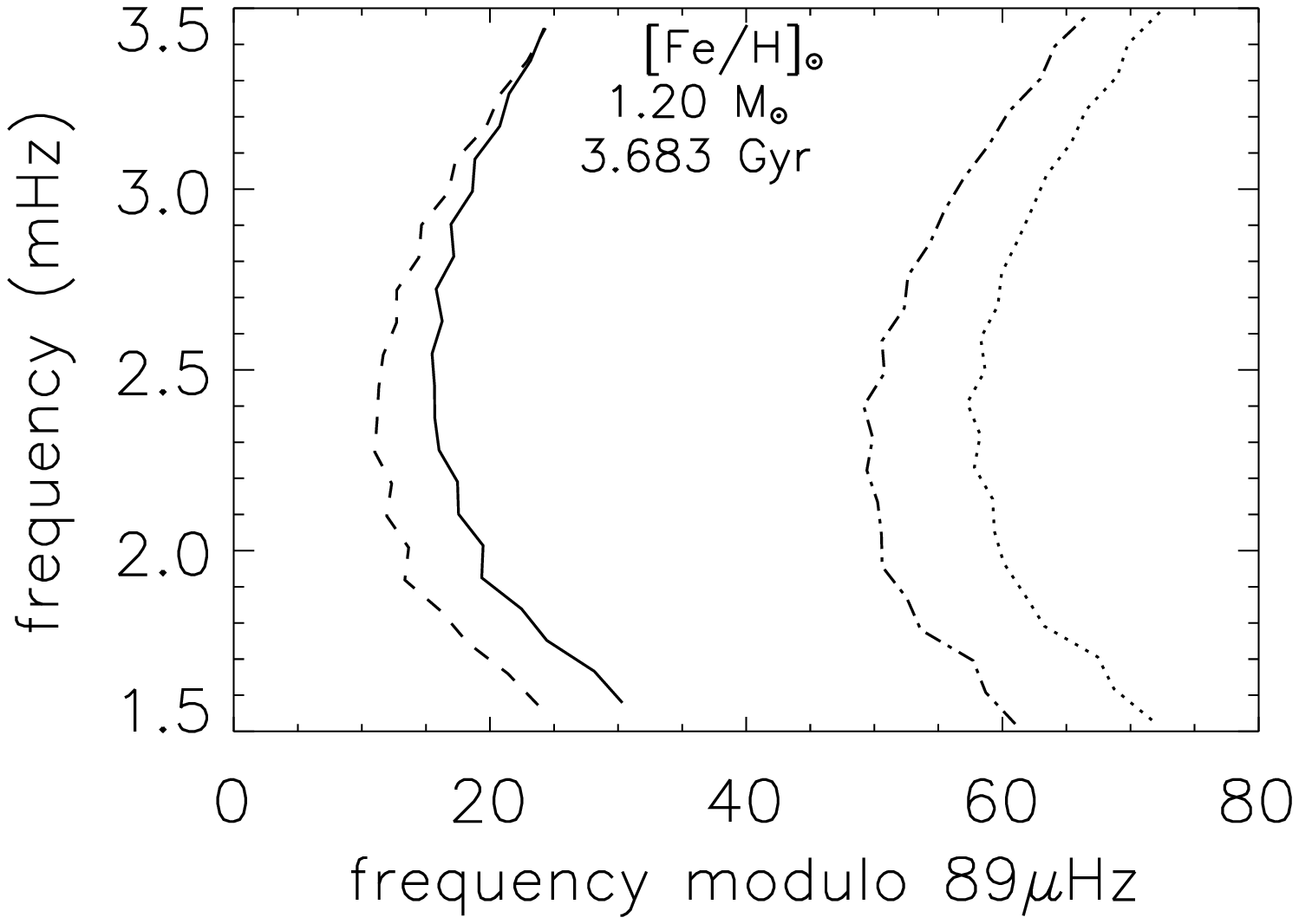}
\includegraphics[angle=0,totalheight=5.5cm,width=8cm]{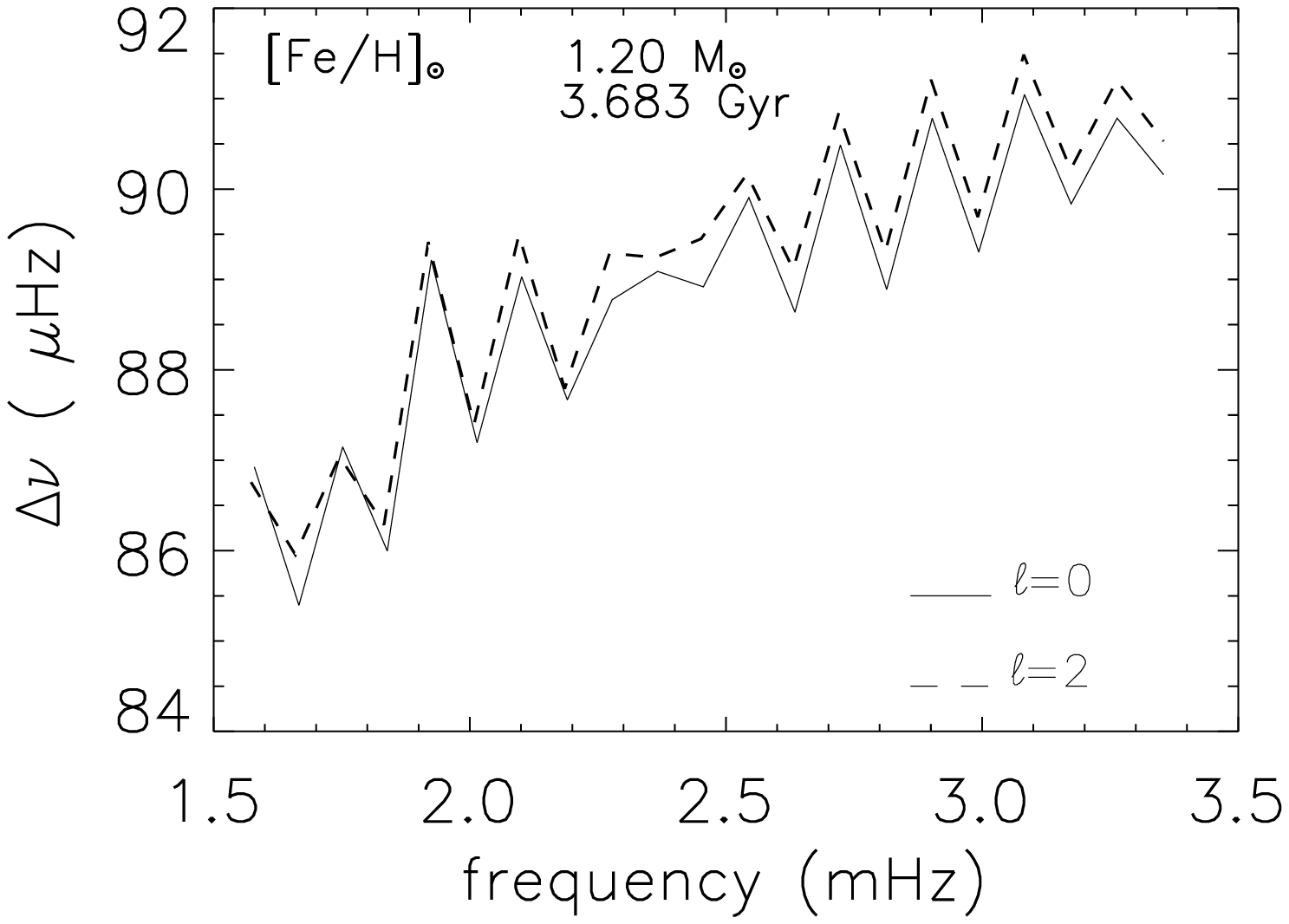}\includegraphics[angle=0,totalheight=5.5cm,width=8cm]{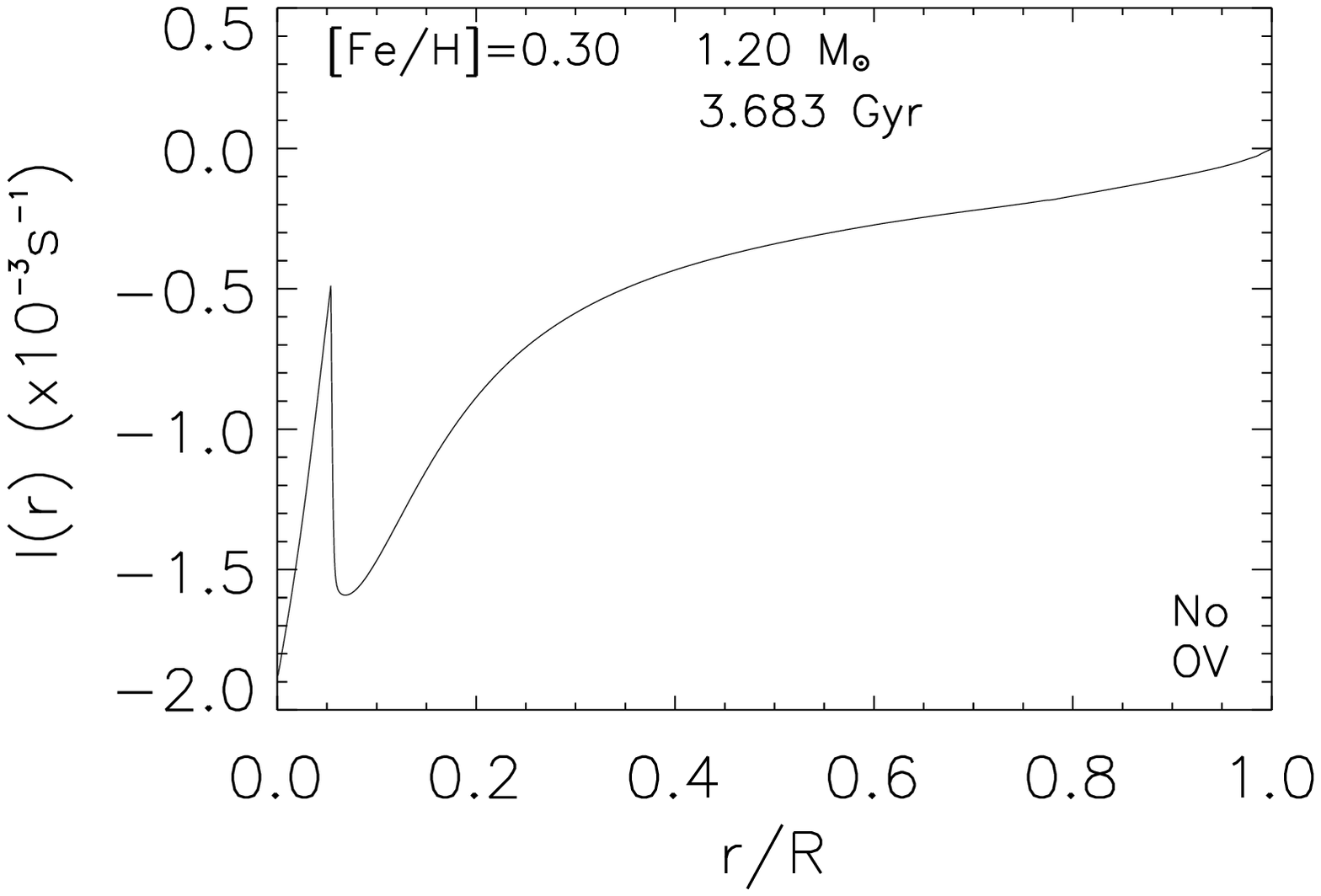}
\end{center}
\caption{Small separations (upper-left panel), echelle diagram (upper-right panel), large separations (lower-left panel), and integral $I(r)$ (lower-right panel) for a model with a solar metallicity, 1.20~\msol  and 3.683~Gyr. It is a main sequence model with a convective core. Here the integral $I(r)$, although rapidly changing near the core, does not change sign. On the other hand, as can be seen in the lower-left graph, the large separations are higher for $\ell=2$ than for $\ell=0$, which is enough to induce a sign change in the small separations. For the echelle diagram, solid lines are for $\ell=0$, dotted lines for $\ell=1$, dashed lines for $\ell=2$, and dotted-dashed lines for $\ell=3$.}
\label{fig3}
\end{figure*}

\begin{figure*}
\begin{center}
\includegraphics[angle=0,totalheight=5.5cm,width=8cm]{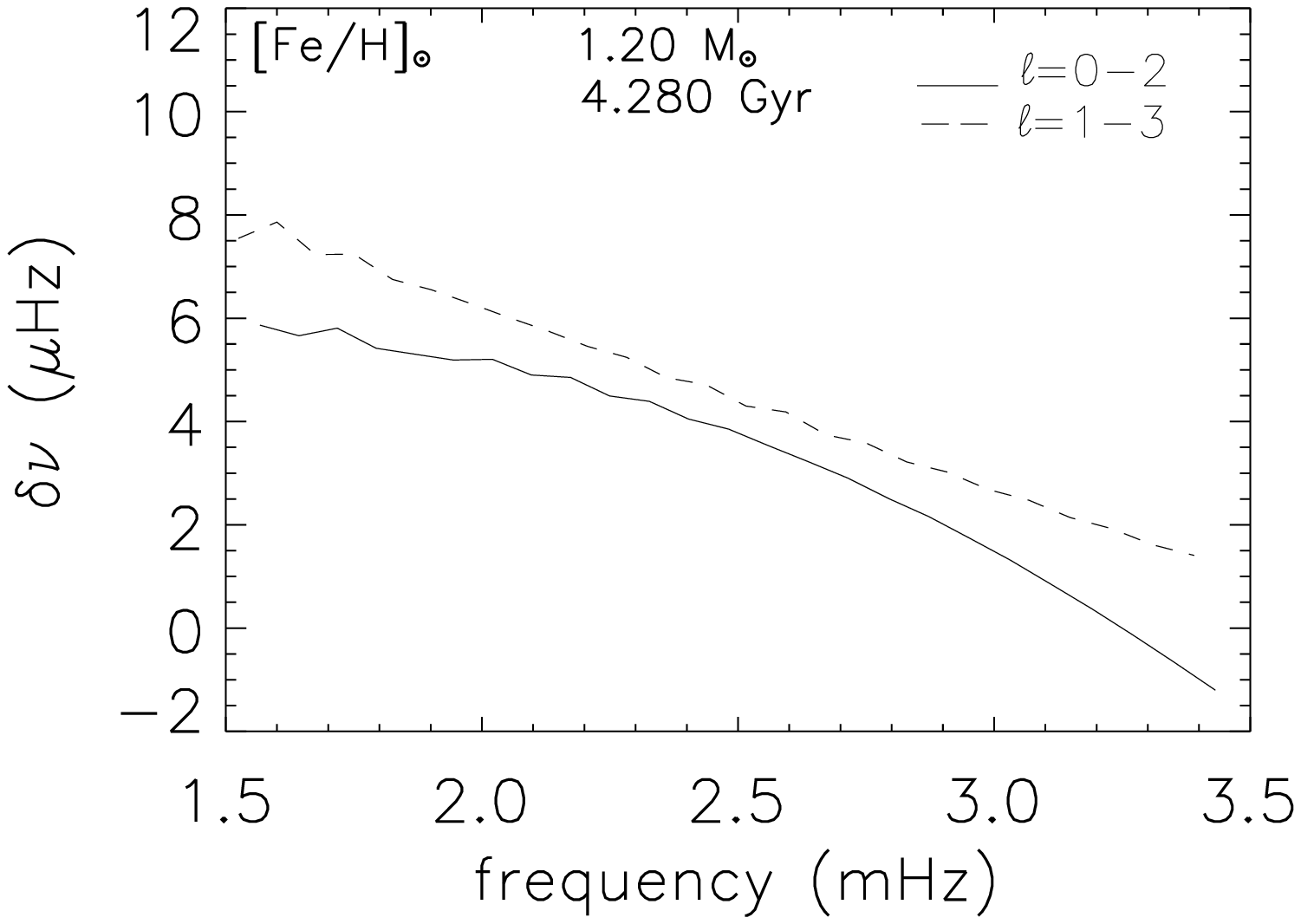}\includegraphics[angle=0,totalheight=5.5cm,width=8cm]{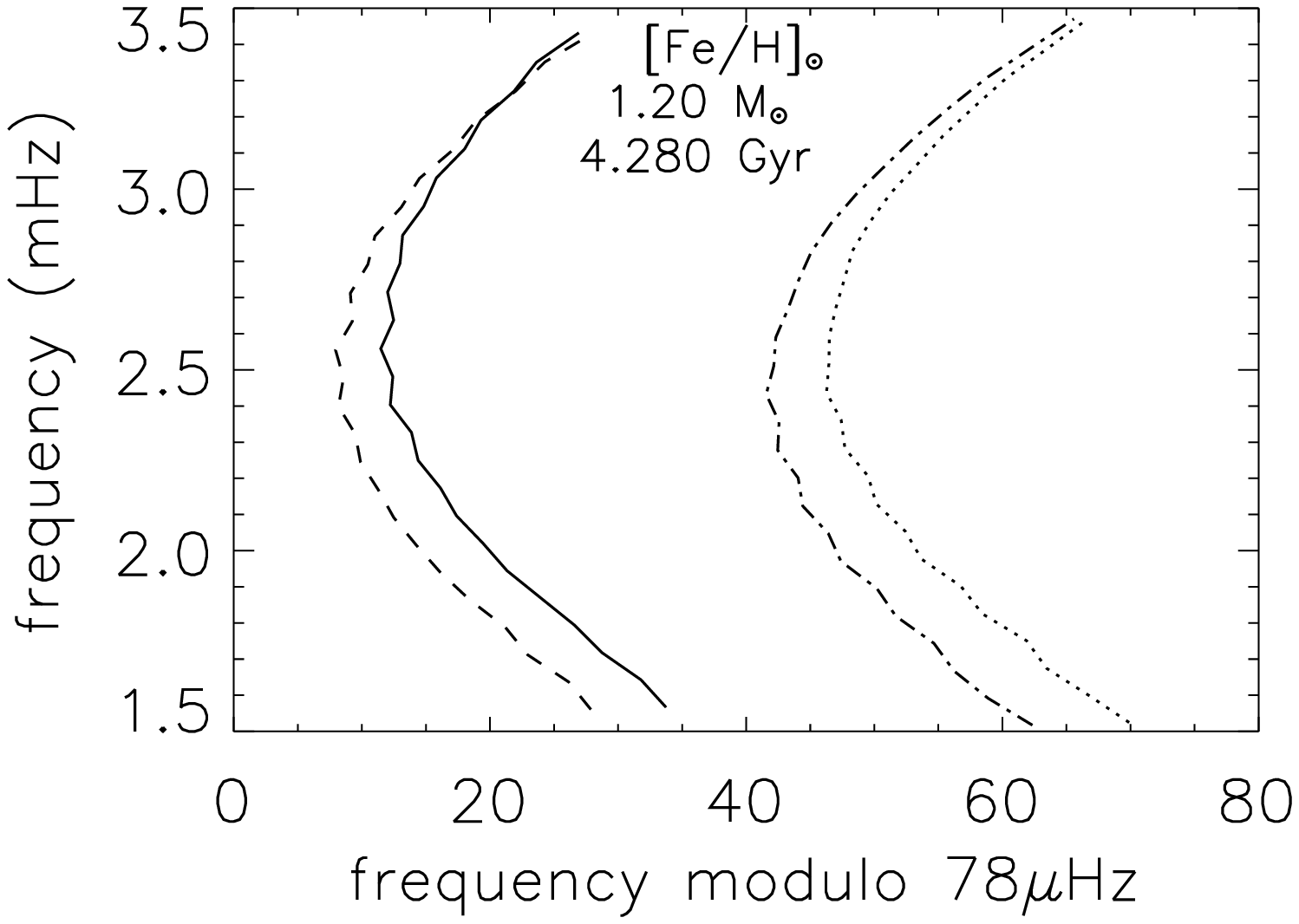}
\includegraphics[angle=0,totalheight=5.5cm,width=8cm]{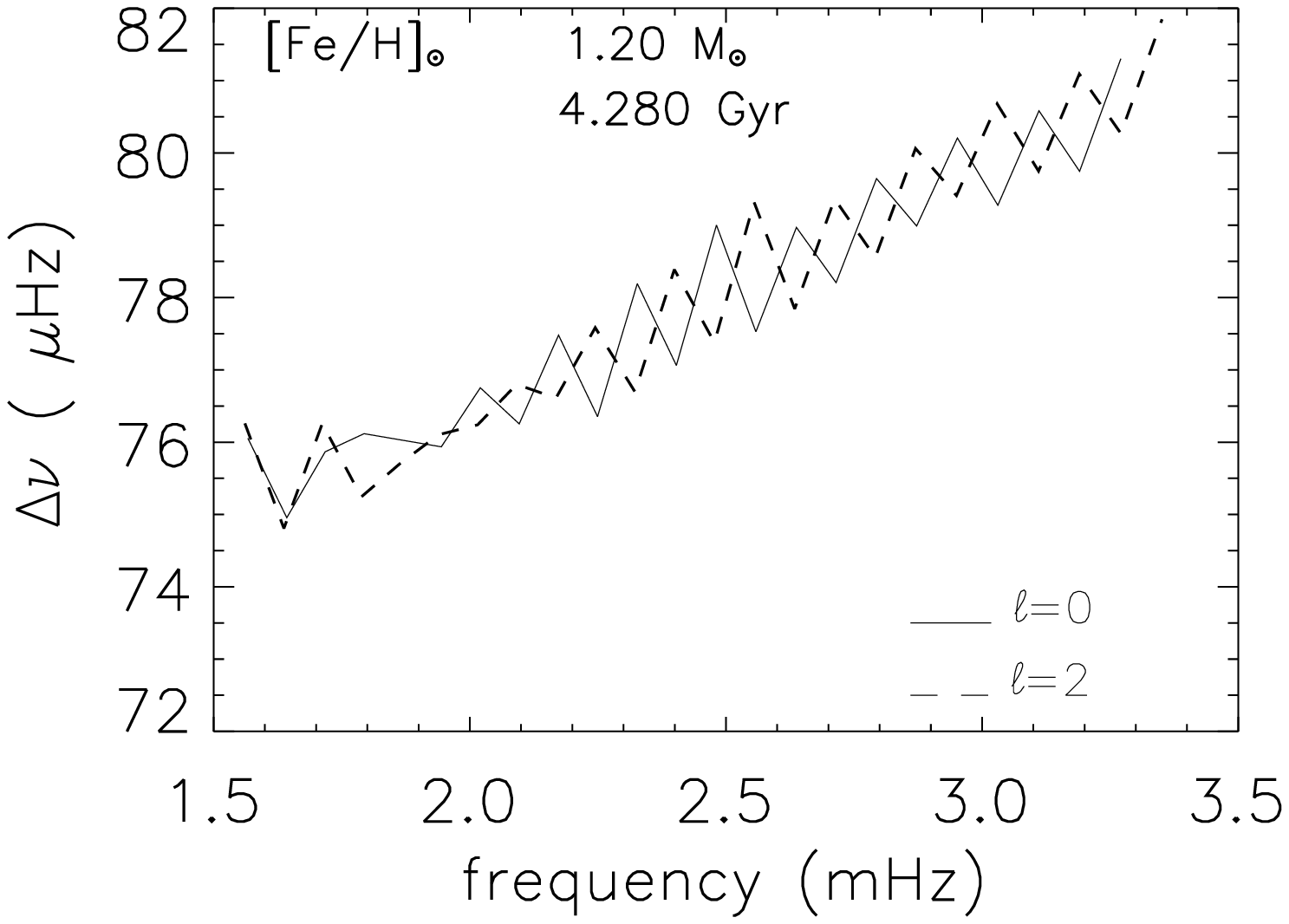}\includegraphics[angle=0,totalheight=5.5cm,width=8cm]{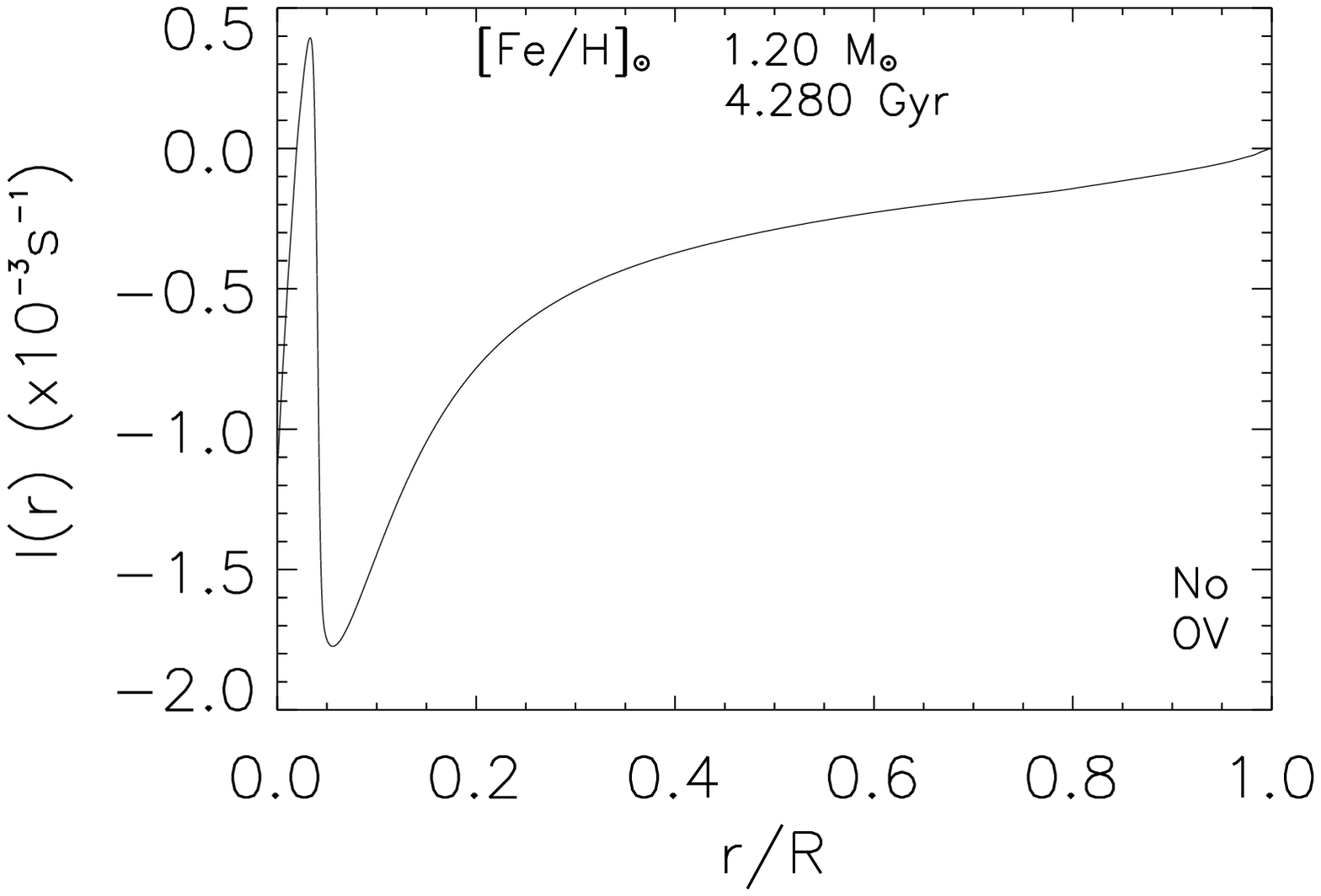}
\end{center}
\caption{Small separations (upper-left panel), echelle diagram (upper-right panel), large separations (lower-left panel), and integral $I(r)$ (lower-right panel) for a model with a solar metallicity, 1.20~\msol  and 4.280~Gyr. It is an evolved model (subgiant branch) with a helium core. Here the integral $I(r)$ changes its sign near the core. This explains the sign change in the small separations that occur when the $\ell=2$ waves reach the core boundary and the crossing point in the echelle diagram. For the echelle diagram, solid lines are for $\ell=0$, dotted lines for $\ell=1$, dashed lines for $\ell=2$, and dotted-dashed lines for $\ell=3$.}
\label{fig4}
\end{figure*}

As discussed in Soriano et al. (\cite{soriano07}), the asymptotic theory has been derived with some assumptions that are no longer valid in the cases we are studying. The most drastic of these assumptions is related to integrals that should be computed between the internal turning point of the waves $r_t$ and the stellar surface, whereas they are replaced by integrations from zero to R. This means that the sound velocity at the turning point is assumed identical to that in the stellar centre. This assumption is completely wrong in stars with convective cores, and it becomes worse and worse as the helium-to-hydrogen ratio increases. A second important assumption is that the large separations for all frequencies and all modes are assumed identical to
\begin{eqnarray}
\Delta\nu_0= \left [ 2 \int_0^R \frac{dr}{c} \right ] ^{-1}.
\label{eqn1}
\end{eqnarray}
This is not correct, as the $\ell \neq 0$ modes do not travel down to the stellar centre. They are trapped at the turning point $r_t$, so that for each mode, integral~(\ref{eqn1}) should be computed from $r_t$ to R, not from 0 to R. As a consequence, the mean large separation for a mode $\ell$ is larger than $\Delta\nu_0$, due to the smaller integral. This effect can be important if the sound velocity decreases strongly in the central regions.

Following Tassoul (\cite{tassoul80}), but relaxing these assumptions, Soriano et al. (\cite{soriano07}) derived the following approximate expressions for the $\ell$~=~0~-~$\ell$~=~2 and for the $\ell$~=~1~-~$\ell$~=~3 small separations:

\begin{eqnarray}
\delta\nu_{02} \simeq (n+\frac{1}{4}+\alpha)(\Delta\nu_{0}-\Delta\nu_{2}) - \frac{6\Delta\nu_{2}}{4\pi^2 \nu_{n-1,2}} \int_{r_t}^R \frac{1}{r} \frac{dc}{dr} dr \\
\delta\nu_{13} \simeq (n+\frac{3}{4}+\alpha)(\Delta\nu_{1}-\Delta\nu_{3}) + \left ( \frac{\Delta\nu_1}{2\pi^2\nu_{n,1}} - \frac{6\Delta\nu_3}{2\pi^2\nu_{n-1,3}} \right ) \int_{r_t}^R \frac{1}{r} \frac{dc}{dr} dr
\end{eqnarray}
where $n$ is the radial order of the modes, $\alpha$ corresponds to a surface phase shift, and $\Delta\nu_0$, $\Delta\nu_1$, $\Delta\nu_2$, and $\Delta\nu_3$ are the large separations computed respectively for the degrees $\ell=0$, $1$, $2$, and $3$. 

From now on, we basically concentrate on the $\delta\nu_{02}$ separations, which are the most relevant for our purpose.
We can see that the integral $I(r)=\int_{r_t}^R \frac{1}{r}\frac{dc}{dr} dr$ plays an important role in the computations of the small separations. Using the real boundary value $r_t$ instead of zero can lead to significant changes in the results. Also the difference between the large separations for the $\ell=0$ and the $\ell=2$ modes, which vanishes in the asymtotic theory, can be important, even if it is small. Indeed this difference is multiplied by the radial order $n$ which can be significant for the considered modes (of order 30). 

We will see below that for some models, the integral $I(r)$, which is generally negative, may become positive at the boundary of the core when the helium over hydrogen ratio is high enough. This is due to the resulting rapid variation of the sound velocity, which is significantly lower in the core than expected from simple extrapolation below the outer layers. In this case, the sign inversion in the integral is sufficient to explain why the small separations become negative (Fig.~\ref{fig4}). The frequency for which this occurs corresponds to that of the $\ell=2$ waves for which the turning point is at the core boundary. 

In other cases, it may happen that the integral value changes abruptly at the core boundary, but not enough to become positive. Then, a slight difference between the large separations for the $\ell=0$ and the $\ell=2$ modes may be enough to lead to the inversion of the sign of the small separations (Fig.~\ref{fig3}). In this case, the frequency where this behaviour occurs may be slightly different from that related to the core boundary.

As a conclusion, that the small separations change sign is related to the core boundary and may be physically understood. However the use of this sign inversion to derive the core size can only be done with precise model computations.

\subsection{Analysis of the small separations along an evolutionary track}

\begin{figure*}
\begin{center}
\includegraphics[angle=0,totalheight=5.5cm,width=8cm]{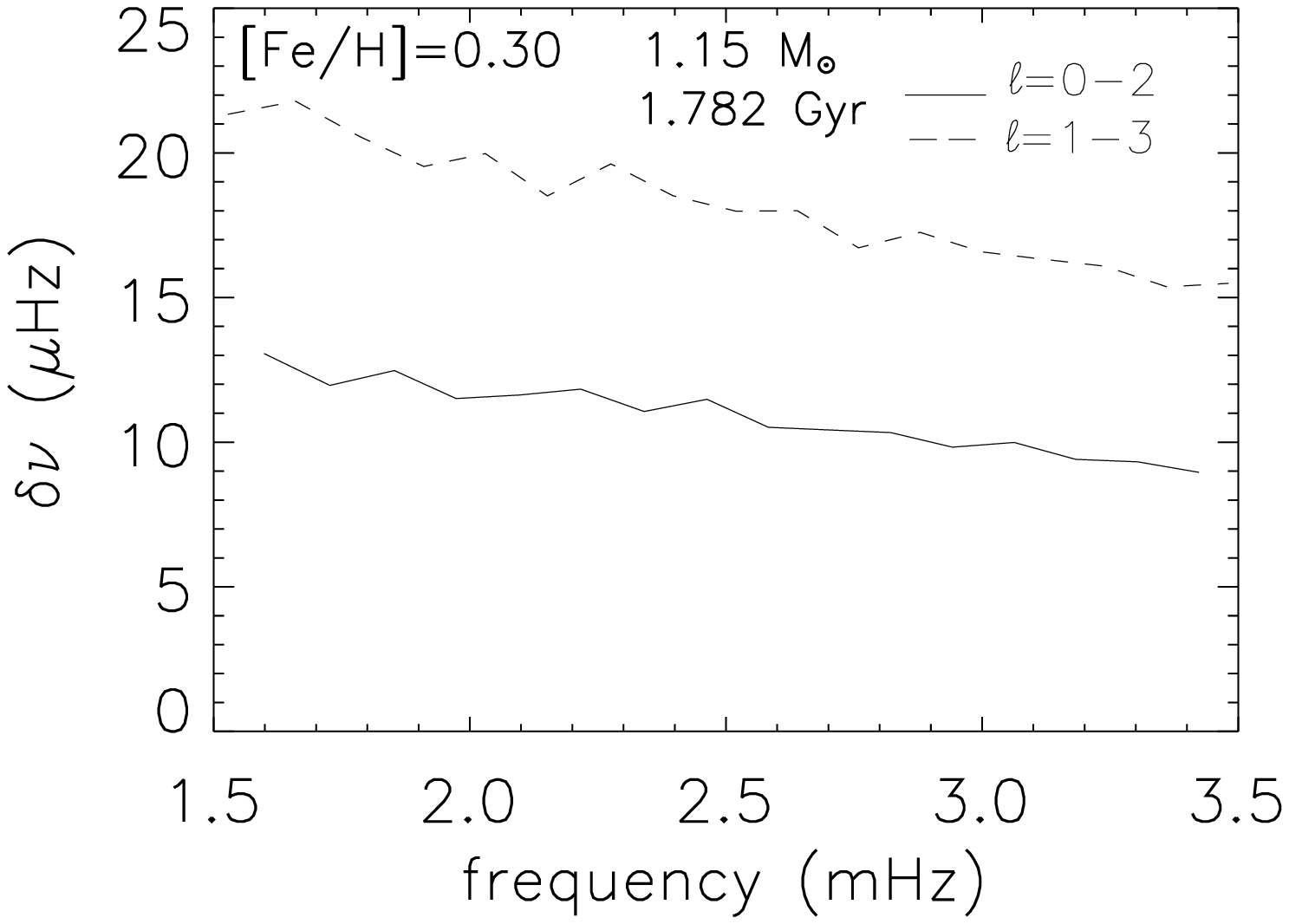}\includegraphics[angle=0,totalheight=5.5cm,width=8cm]{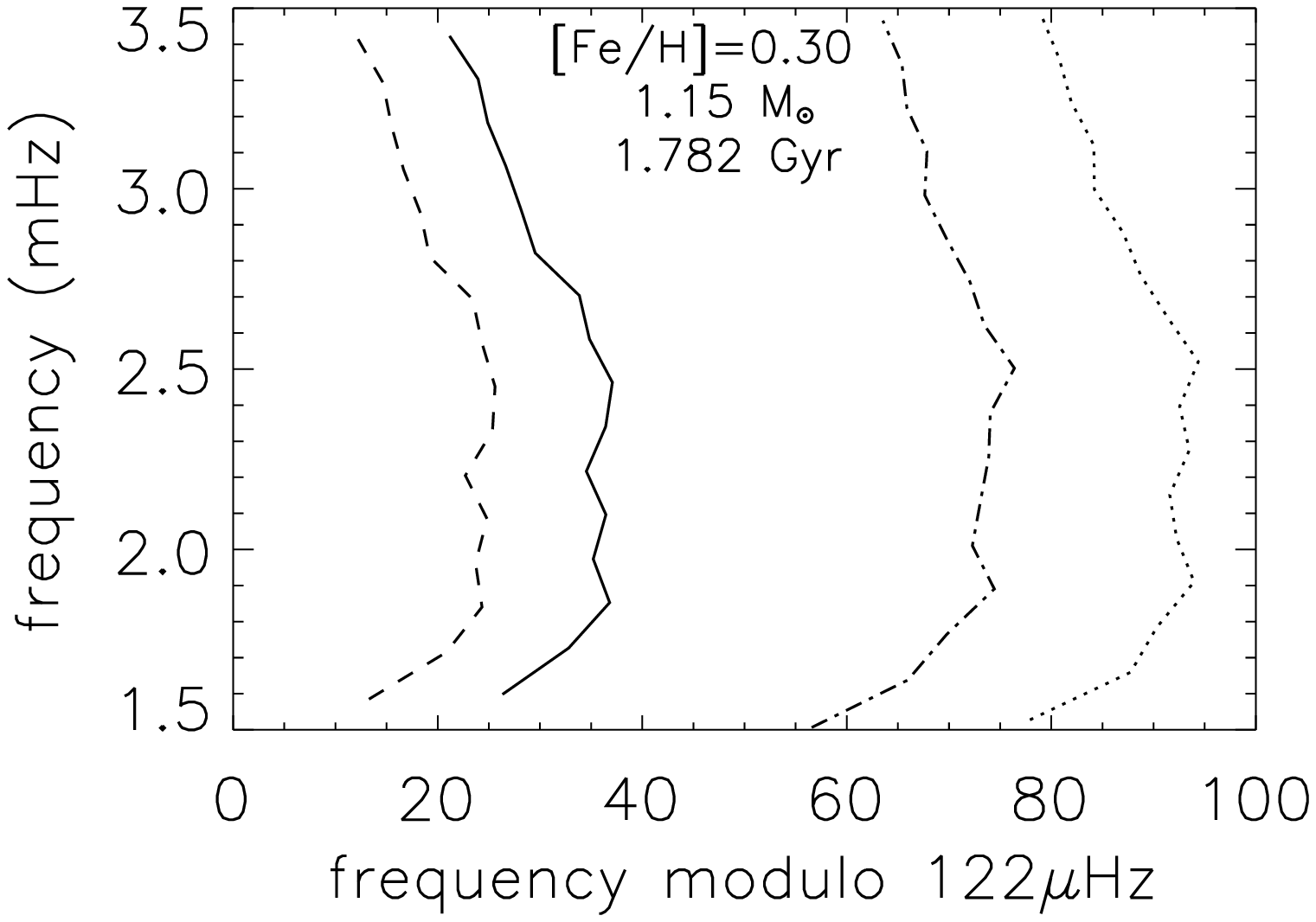}
\includegraphics[angle=0,totalheight=5.5cm,width=8cm]{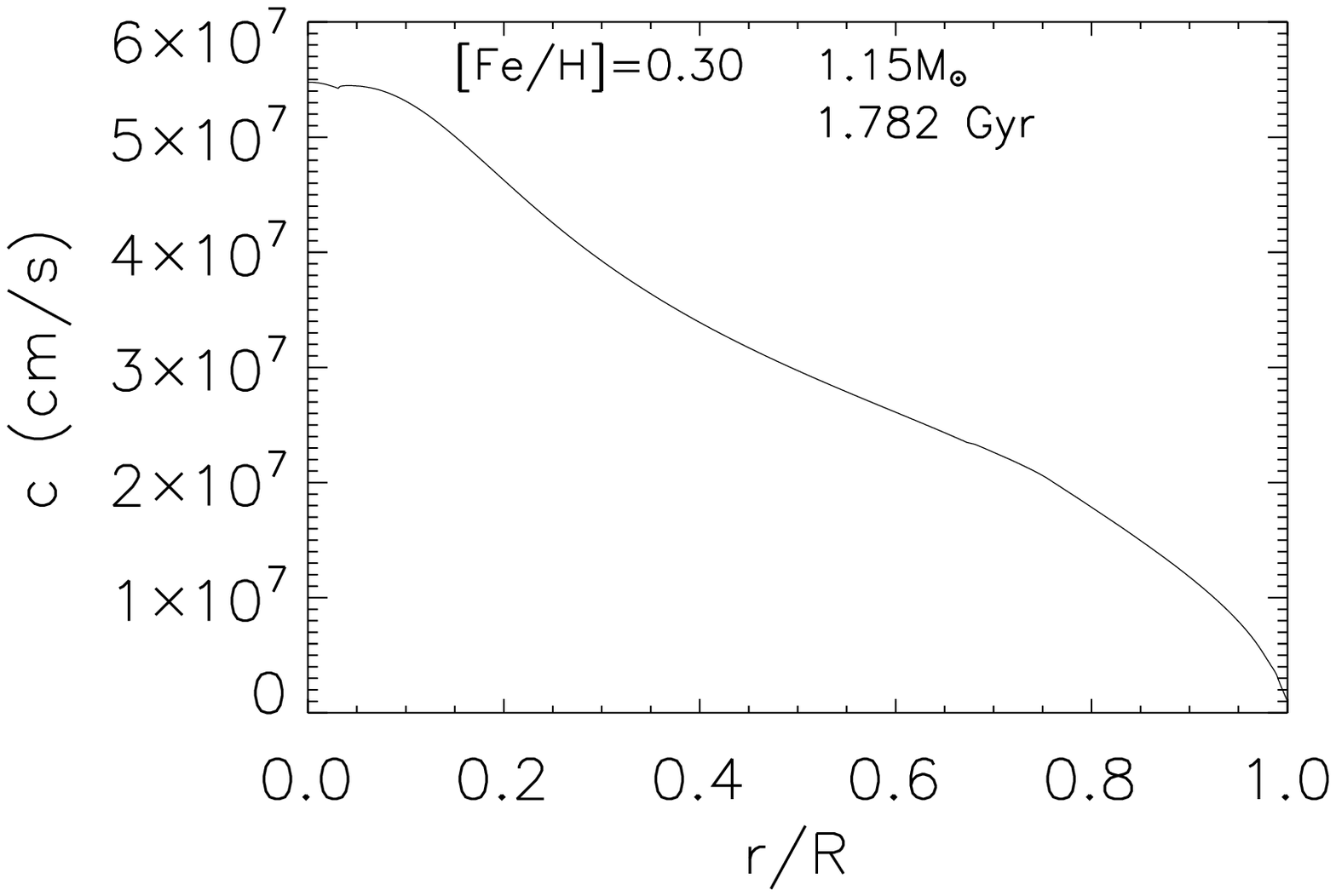}\includegraphics[angle=0,totalheight=5.5cm,width=8cm]{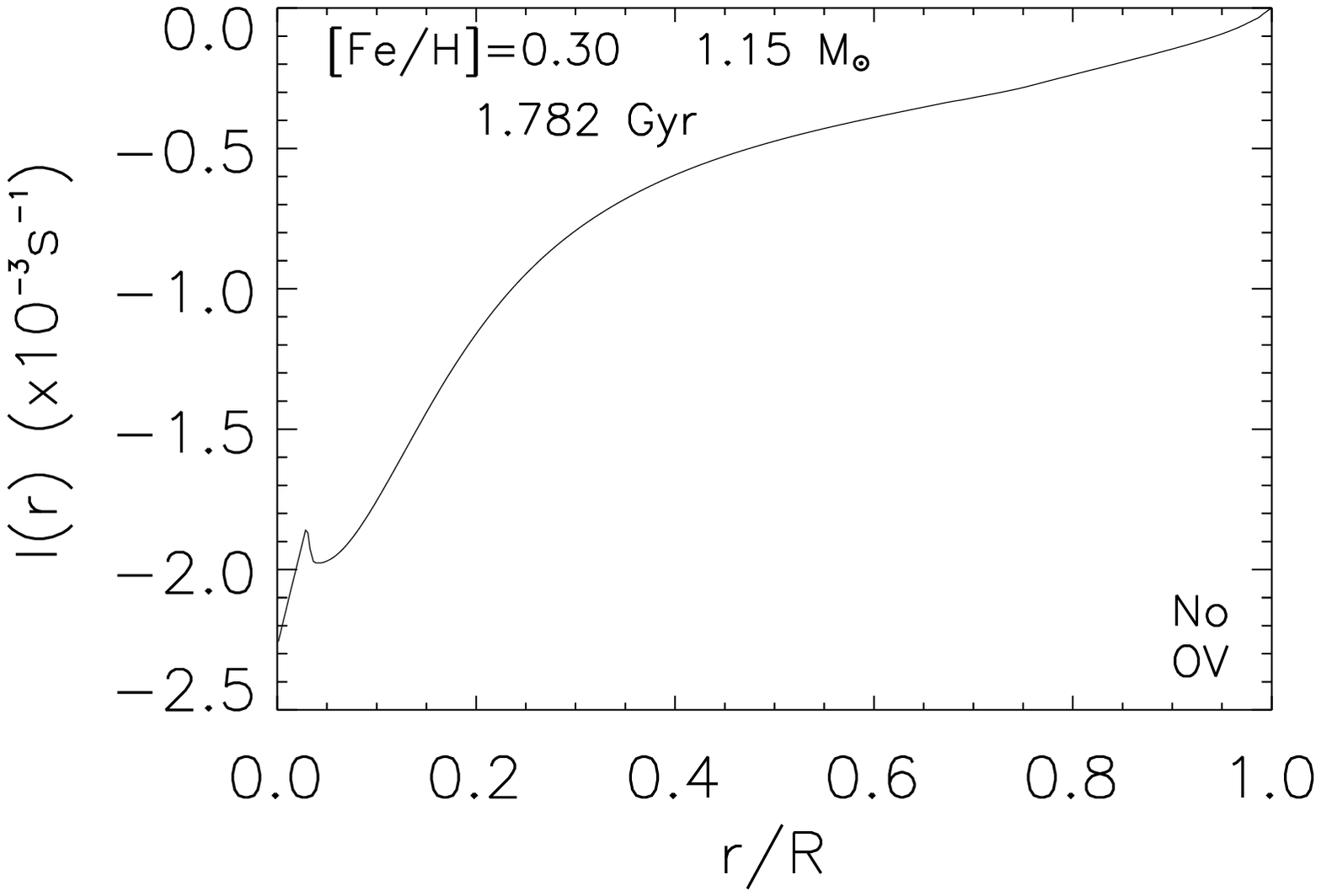}
\end{center}
\caption{Small separations (upper-left panel), echelle diagram (upper-right panel), sound speed profile (lower-left panel), and integral $I(r)=\int_{r_t}^R \frac{1}{r} \frac{dc}{dr} dr$ (lower-right panel) for an overmetallic model ([Fe/H]=0.30) of 1.15 \msol and 1.782 Gyr, without overshooting. For the echelle diagram, solid lines are for $\ell=0$, dotted lines for $\ell=1$, dashed lines for $\ell=2$, and dotted-dashed lines for $\ell=3$.}
\label{fig5}
\end{figure*}

\begin{figure*}
\begin{center}
\includegraphics[angle=0,totalheight=5.5cm,width=8cm]{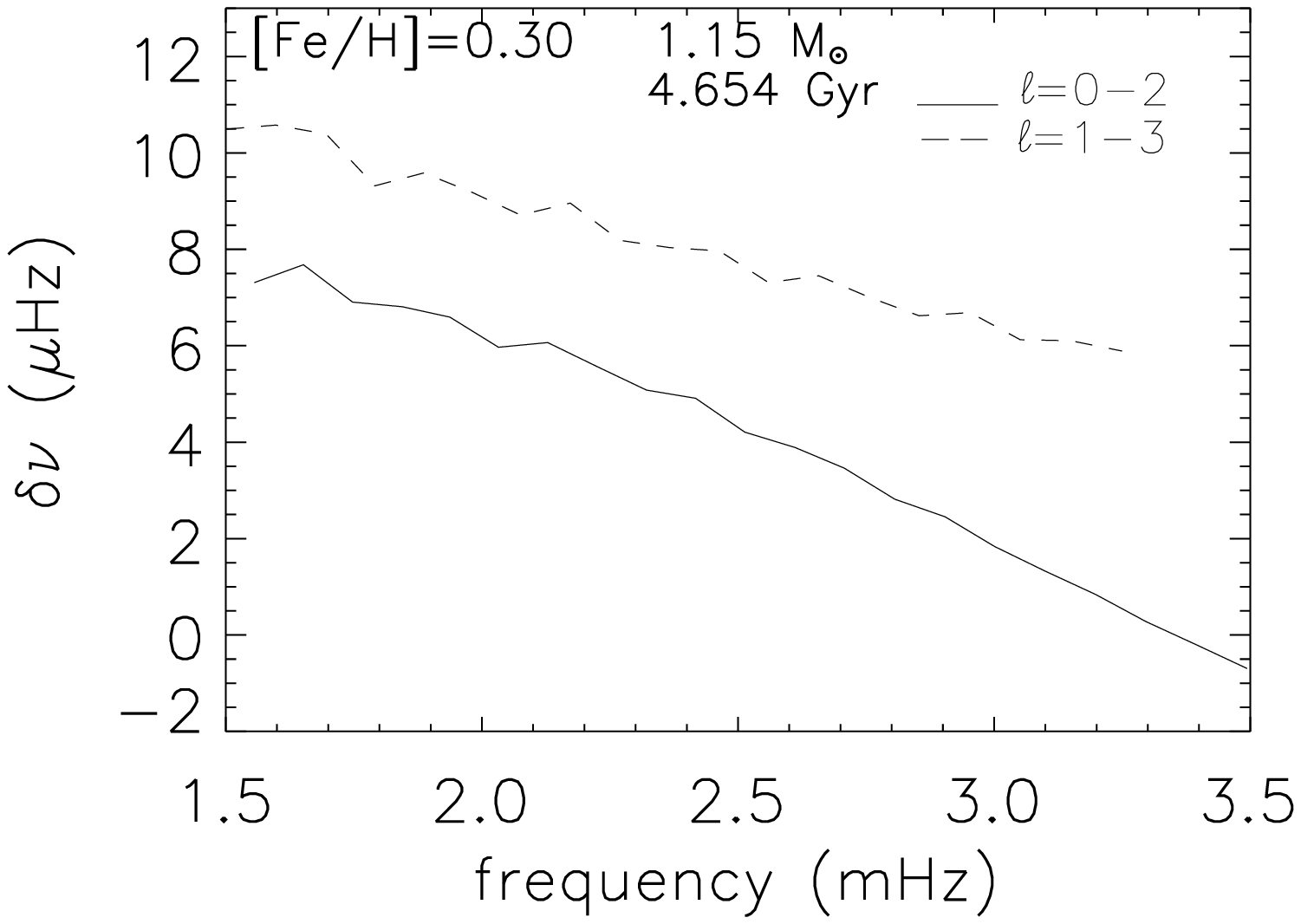}\includegraphics[angle=0,totalheight=5.5cm,width=8cm]{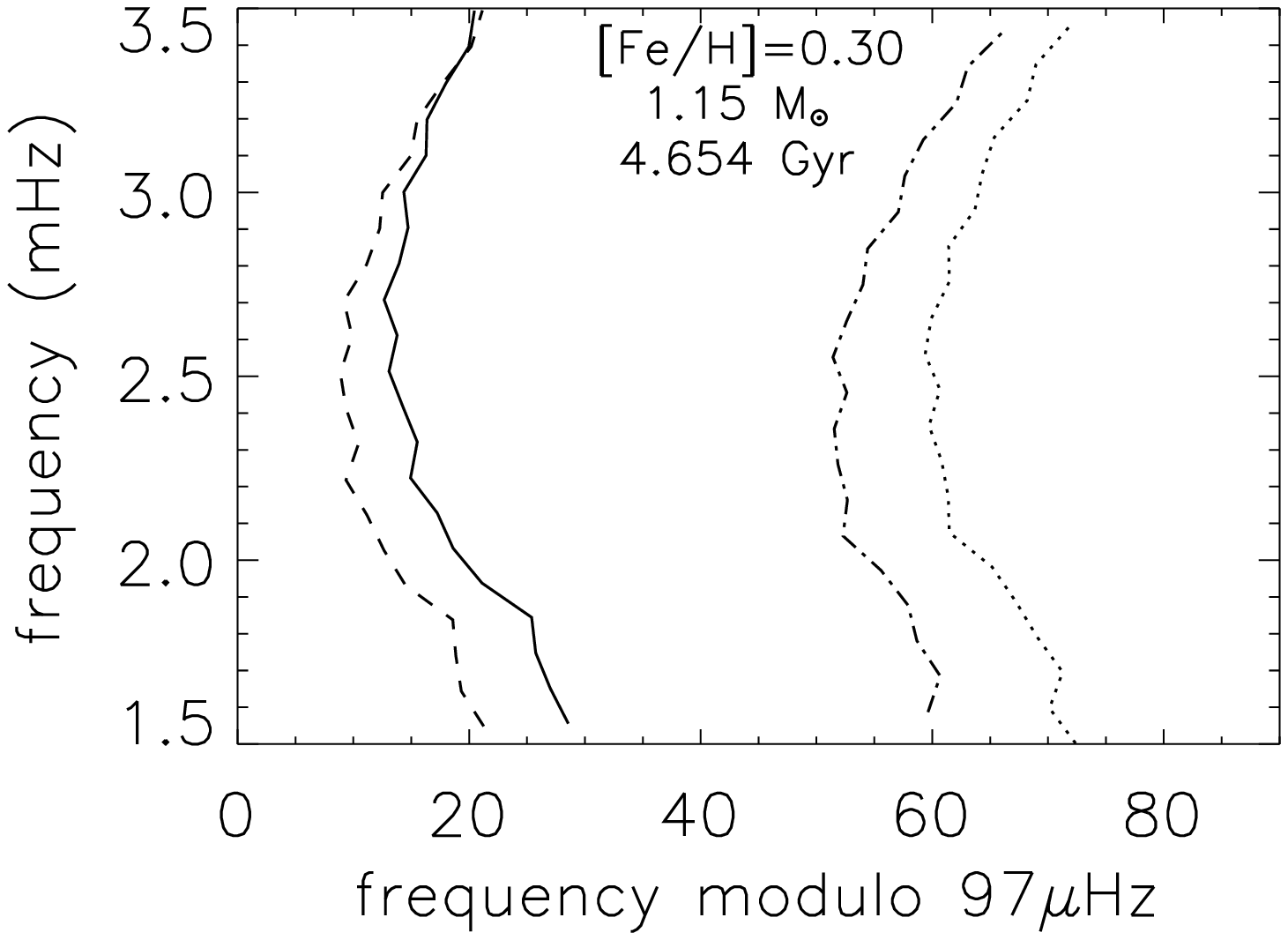}
\includegraphics[angle=0,totalheight=5.5cm,width=8cm]{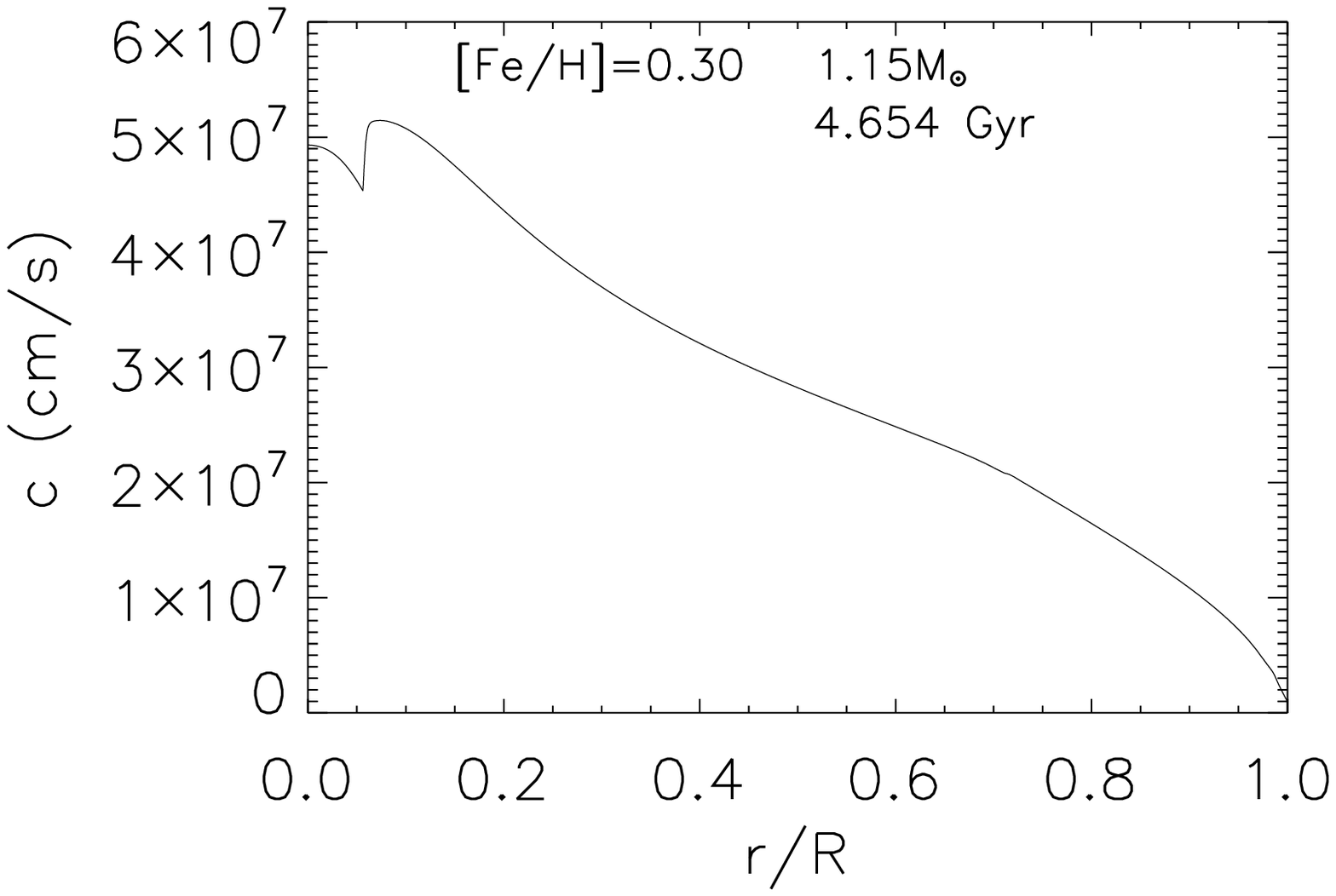}\includegraphics[angle=0,totalheight=5.5cm,width=8cm]{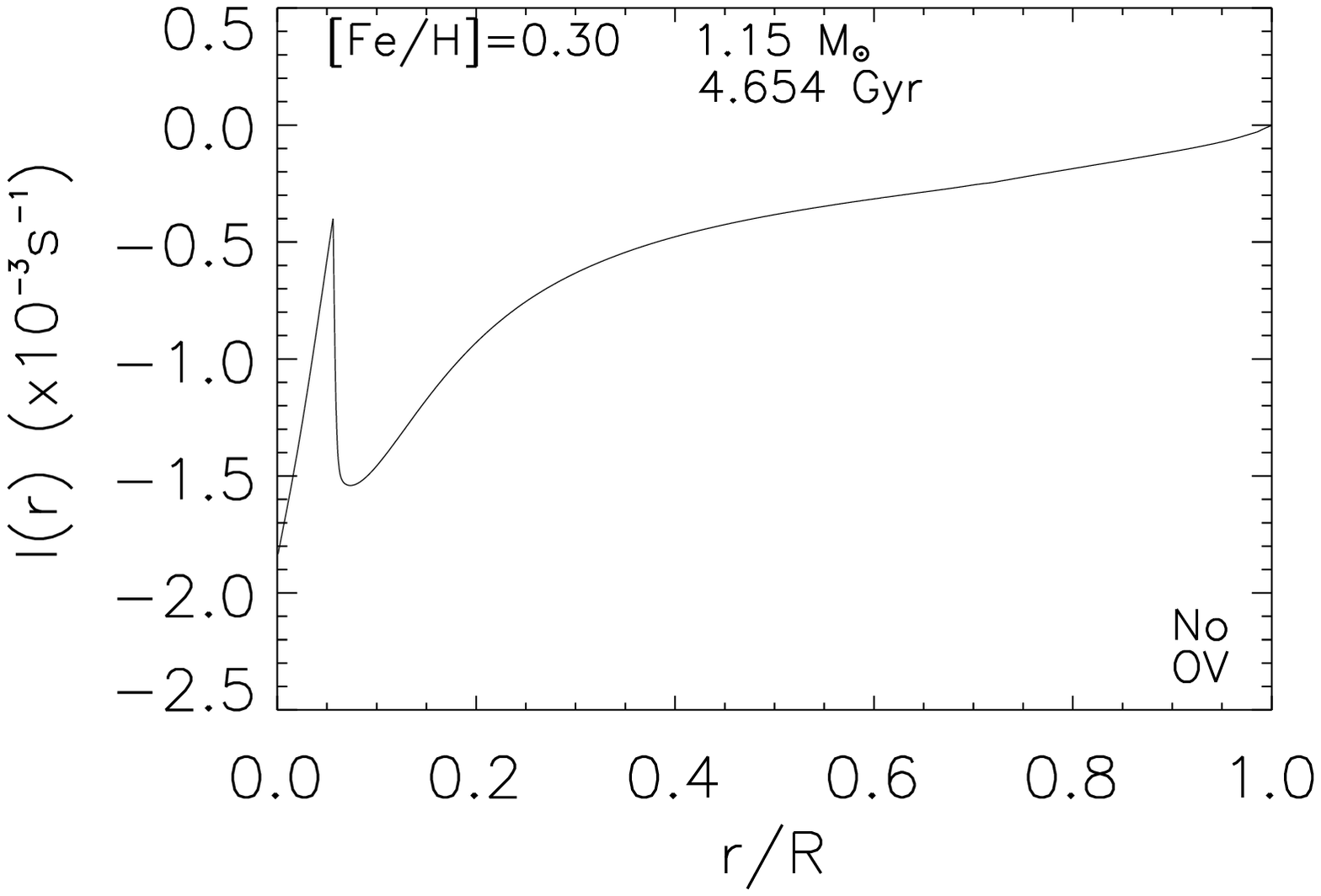}
\end{center}
\caption{Small separations (upper-left panel), echelle diagram (upper-right panel), sound speed profile (lower-left panel), and integral $I(r)=\int_{r_t}^R \frac{1}{r} \frac{dc}{dr} dr$ (lower-right panel) for an overmetallic model ([Fe/H]=0.30) of 1.15 \msol and 4.654 Gyr, without overshooting. It is the first model for which we obtain negative small separations below 3.5 mHz. This model has a convective core. There is an important discontinuity in its sound speed profile at the boundary of the core. For the echelle diagram, solid lines are for $\ell=0$, dotted lines for $\ell=1$, dashed lines for $\ell=2$, and dotted-dashed lines for $\ell=3$.}
\label{fig6}
\end{figure*}

\begin{figure*}
\begin{center}
\includegraphics[angle=0,totalheight=5.5cm,width=8cm]{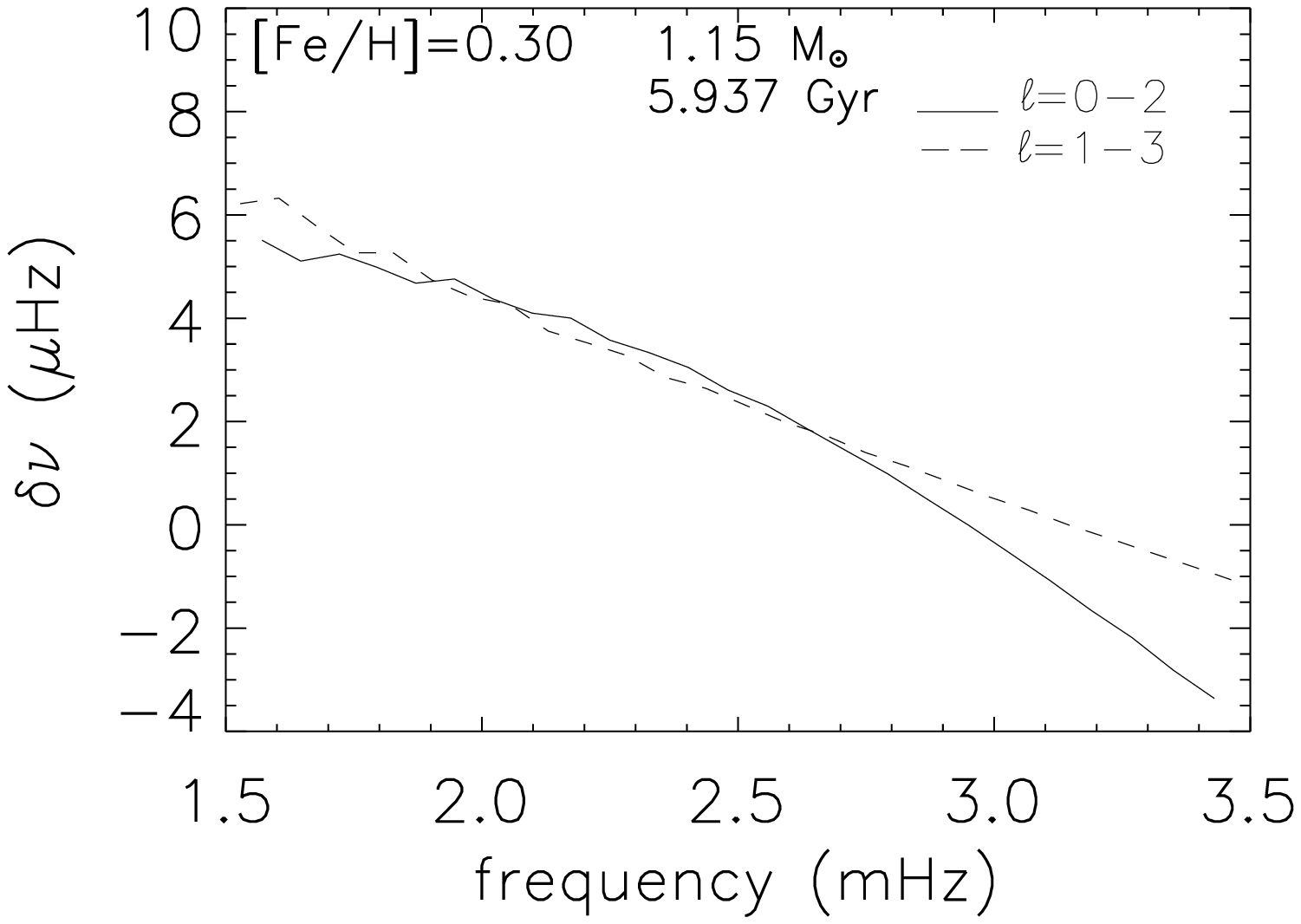}\includegraphics[angle=0,totalheight=5.5cm,width=8cm]{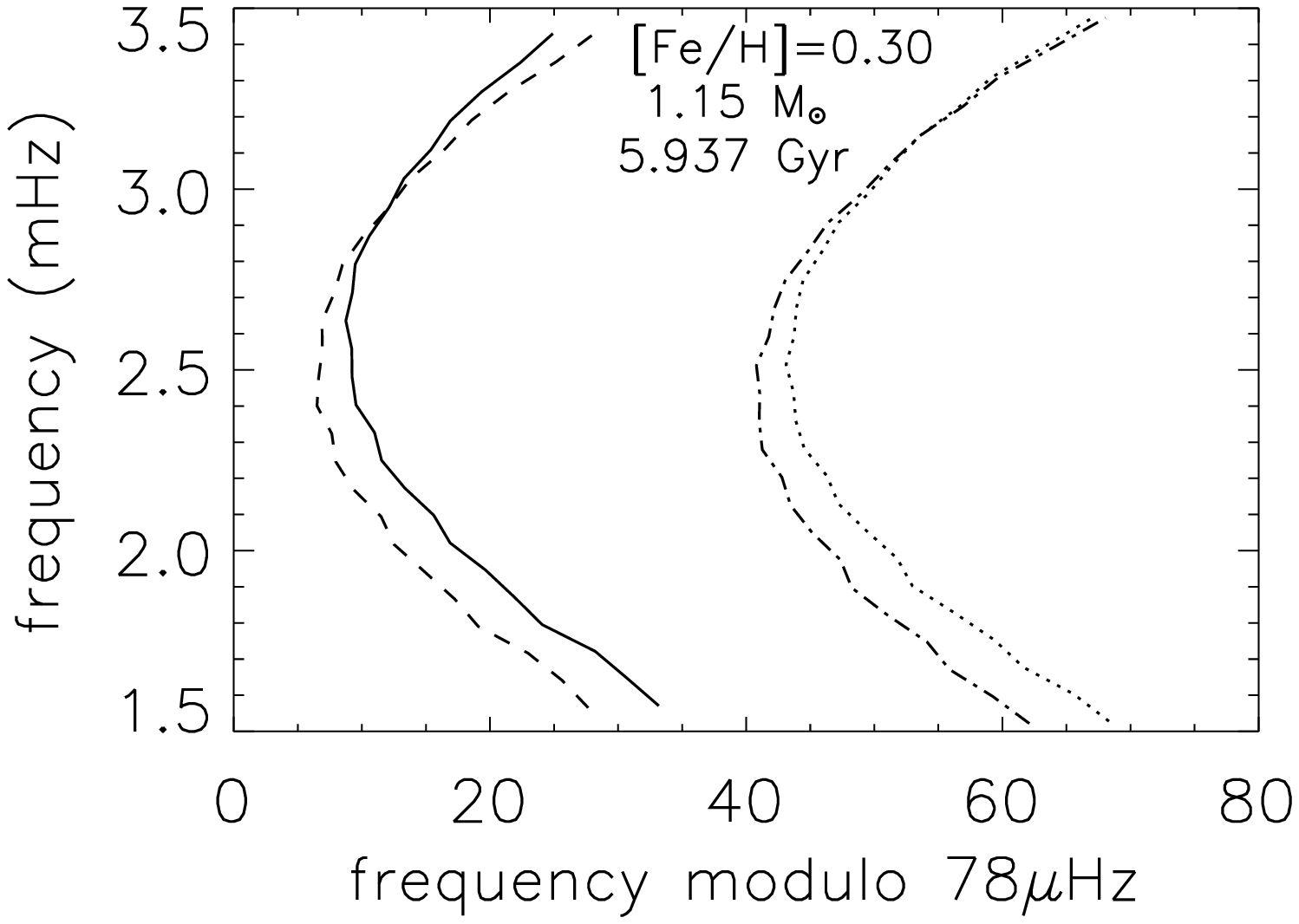}
\includegraphics[angle=0,totalheight=5.5cm,width=8cm]{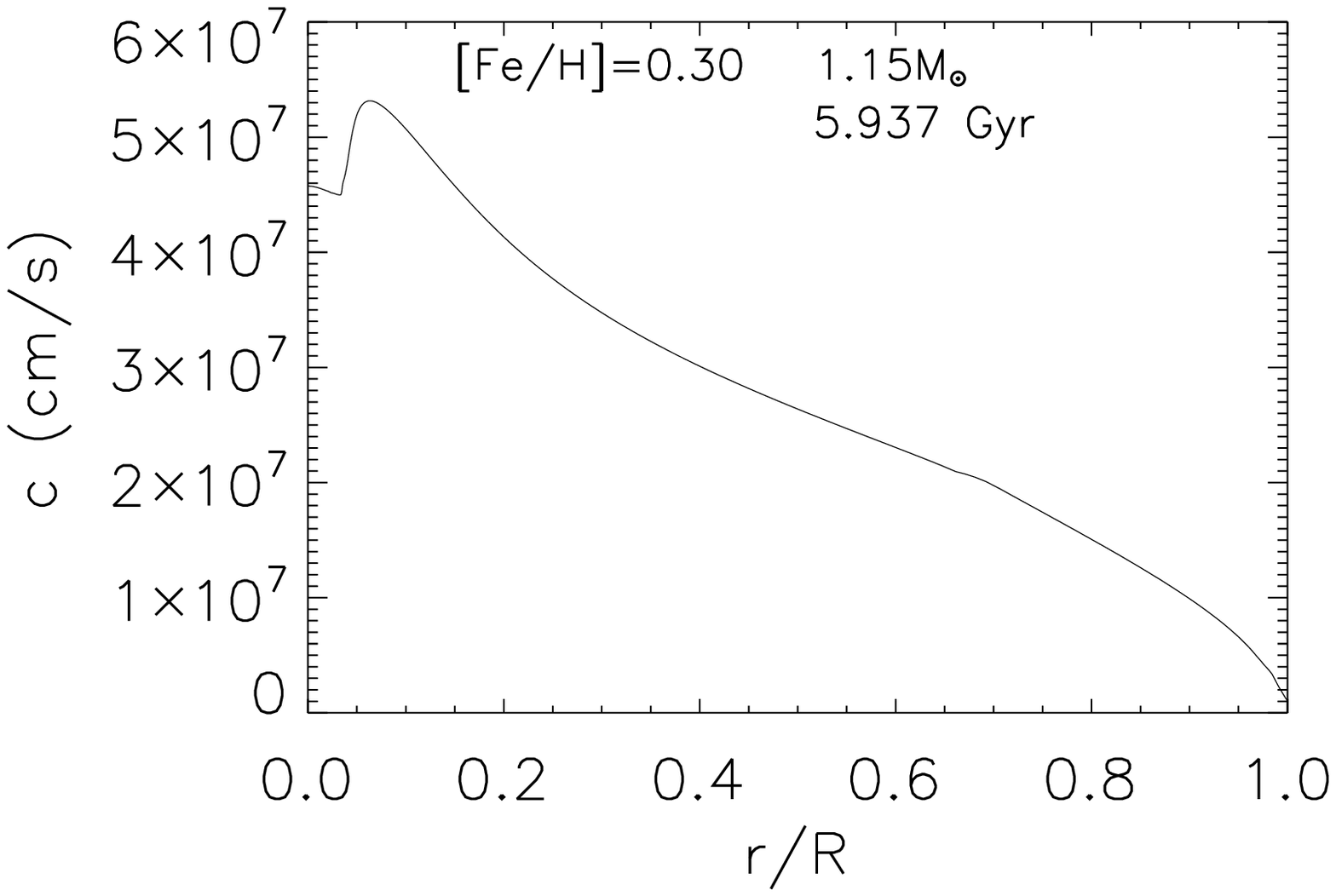}\includegraphics[angle=0,totalheight=5.5cm,width=8cm]{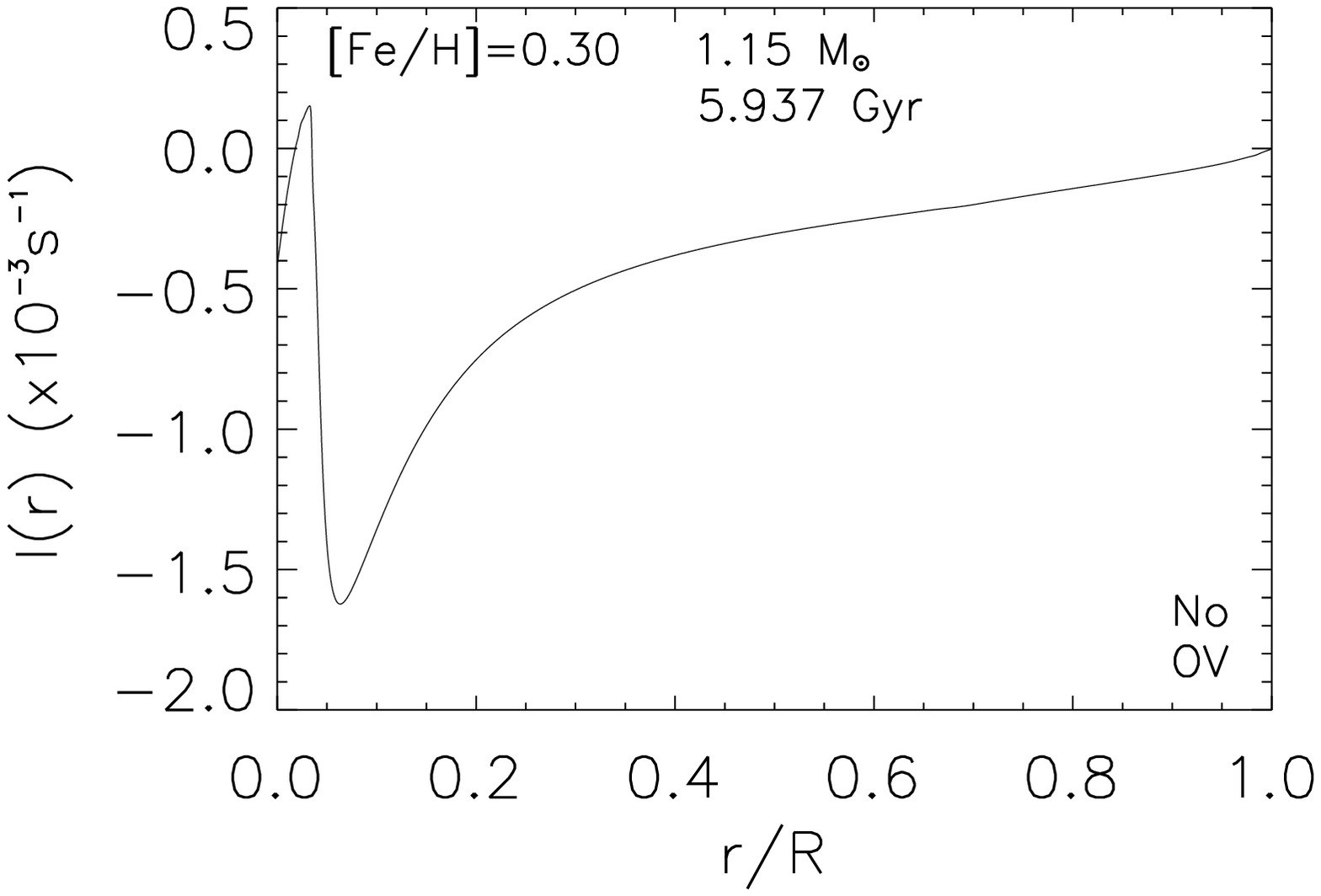}
\end{center}
\caption{Small separations (upper-left panel), echelle diagram (upper-right panel), sound speed profile (lower-left panel), and integral $I(r)=\int_{r_t}^R \frac{1}{r} \frac{dc}{dr} dr$ (lower-right panel) for an overmetallic model ([Fe/H]=0.30) of 1.15~\msol and 5.937 Gyr without overshooting. This model is on the subgiant branch with a helium-rich core. Here the integral $I(r)$ changes sign near the core. For the echelle diagram, solid lines are for $\ell=0$, dotted lines for $\ell=1$, dashed lines for $\ell=2$, and dotted-dashed lines for $\ell=3$.}
\label{fig7}
\end{figure*}

Figures~\ref{fig5} to \ref{fig7} present the internal parameters for three models along an evolutionary track of 1.15 \msol with [Fe/H]=0.30, without overshooting. The first model lies on the main sequence, in a situation where the small separations are always positive in the considered frequency range (1.782 Gyr, Fig.~\ref{fig5}). The second model is more evolved, and the small separations for the degrees $\ell=0$~-~$\ell=2$ become negative at a frequency of 3.4 mHz (4.654 Gyr, Fig.\ref{fig6}). It corresponds to what we have called the ``transition model''. The third model is still more evolved on the subgiant branch (5.937 Gyr, Fig.~\ref{fig7}).

In each figure, four different graphs are presented: the small separations (upper-left panel), the echelle diagram (upper-right panel), the sound speed profile (lower-left panel), and the integral $I(r)$ (lower-right panel).

At the beginning of the main sequence (model presented in Fig.\ref{fig5}), the convective core of the star is still small ($R_{cc}/R=0.03$). The central helium abundance is low and there is no discontinuity in the sound speed profile at the centre of the star. The integral $I(r)$ is negative in the whole star. In that case, the small separations $\delta\nu_{02}$ are positive for the full range of frequencies considered, and there is no crossing point in the echelle diagram.

When the age of the star increases, the values of the small separations decrease. The convective core develops, and there is more helium in the centre of the star. A discontinuity in the chemical composition and in the sound speed profile appears. The model shown in Fig.~\ref{fig6} is the first model for which we found negative small separations below 3.5 mHz. We call it the ``transition model'' and show its position as a cross on the corresponding track in Fig.~\ref{fig1}. The convective core is larger ($r_{cc}/R=0.06$) than for the previous model. We can see the  discontinuity in the sound speed profile (Fig.~\ref{fig6}, upper-left panel). As a consequence, the integral $I(r)$ changes drastically between $r=0$ and $r=r_t=R_{cc}$. Here it does not change its sign but, as discussed before, its variations are large enough to lead to negative small separations. The lines $\ell=0$ and $\ell=2$ cross in the echelle diagram. 

In the third model presented in Fig.~\ref{fig7}, the convective core has disappeared, leaving a helium core. There is still a sharp discontinuity in the sound speed profile at the edge of the core (Fig.~\ref{fig7}, lower-left panel), and the small separations change sign at a lower frequency than for the previous model (Fig.~\ref{fig7}, upper-left panel).

The frequencies at which the small separations become negative go on decreasing while the star evolves along the subgiant branch, until the stellar structure begins to change deeply, with an increase in the outer convective layers and a shrinking of the core. A discussion of this evolutionary stage is beyond the scope of the present paper and will be given elsewhere. Here we concentrate on the main sequence and early subgiant phases.

\subsection{Influence of the stellar mass}

Tables~\ref{tab1} to \ref{tab6} present the parameters for the ``transition models'' for various physical assumptions. Let us recall the meaning of these models. As discussed in the previous section, the frequencies for which the small separations $\delta\nu_{02}$ change sign decrease with increasing age along an evolutionary track. The transition model is arbitrarily defined on each track as the one for which this characteristic frequency is about 3.5 mHz. The interest of comparing such specific models is that it helps in analysing the influence of parameters like stellar mass or chemical composition on the behaviour of the small separations.

In each table, the results are presented for the same five masses, between 1.05 and 1.25 \msol. We can see that the ages of the transition models decrease for increasing masses. This is not only due to the more rapid evolution time scales, as the corresponding helium mass fractions also decrease with increasing mass. On the other hand, we can see that the total radii of the models, as well as the radii of their convective or helium cores, do not show strong variations with mass. 

These results are quite understandable if we remember the reason for the appearance of the negative small separations. As discussed previously, they generally appear for frequencies for which the turning point of the $\ell=2$ waves corresponds to the edge of the core. The relation between the frequencies of the waves and their turning points is given by (see Soriano et al. \cite{soriano07})
\begin{eqnarray}
2\pi\nu=\frac{c(r)}{r}\sqrt{\ell(\ell+1)},
\end{eqnarray}
where $\nu$ is the frequency of the considered mode, $r$ the radius, and $c(r)$ the sound speeed.
So that, for the $\ell=2$ waves, the critical frequency for which the turning point is at the core edge is approximately 
\begin{eqnarray}
\nu_c \simeq 0.4 \frac{c(R_{cc})}{R_{cc}}. 
\end{eqnarray}

Although $c(R_{cc})$ may slightly vary from model to model, it is not surprising to find that fixed values of $\nu_c$ corresponding to models with similar values of $R_{cc}$. As a consequence, this behaviour appears earlier for more massive stars and for models with similar sizes, but different luminosities, effective temperatures, and gravities.

\subsection{Influence of the chemical composition.}

Tables~\ref{tab1} and \ref{tab4} present the parameters of the transition models computed with the solar composition (here we use the Grevesse \& Noels \cite{grevesse93} composition, compatible with helioseismology). As will be discussed below, Table~\ref{tab4} includes overshooting, which is not the case for Table~\ref{tab1}. Tables~\ref{tab2} and \ref{tab5} correspond to models computed with a larger abundance of metals ([Fe/H] = 0.30) and a helium value following the metallicity-helium relation obtained for the chemical evolution of galaxies (Izotov \& Thuan \cite{izotov04}). Finally Tables~\ref{tab3} and \ref{tab6} correspond to models with the same overmetallicity but with a solar helium abundance \ysol.

The strongest effect of overmetallicity is to move the evolutionary tracks to lower effective temperatures. This is true for both helium values. In the case of the large helium mass fraction, the ages of the overmetallic transition models are similar to the case of solar values; however, this corresponds to different stages of evolution. For example, the models of 1.15 \msol~~presented in Tables~\ref{tab1} and \ref{tab2} have similar ages, but the one with overmetallicity and large helium abundance (Table~\ref{tab2}) is still on the main sequence with a convective core, whereas the one with solar abundances (Table~\ref{tab1}) is already on the subgiant branch, with a helium core but no convective core anymore. As can be checked in the tables, their luminosities, effective temperatures, and log~$g$ are consequently quite different.

For the case of overmetallic models with a solar helium abundance, the ages of the transition models are higher than for models with solar metallicity. As for the solar case, the less massive models are already on the subgiant branch with no convective core left, but their luminosities and effective temperatures are lower.

\subsection{Influence of overshooting}

\begin{figure*}
\begin{center}
\includegraphics[angle=0,totalheight=5.5cm,width=8cm]{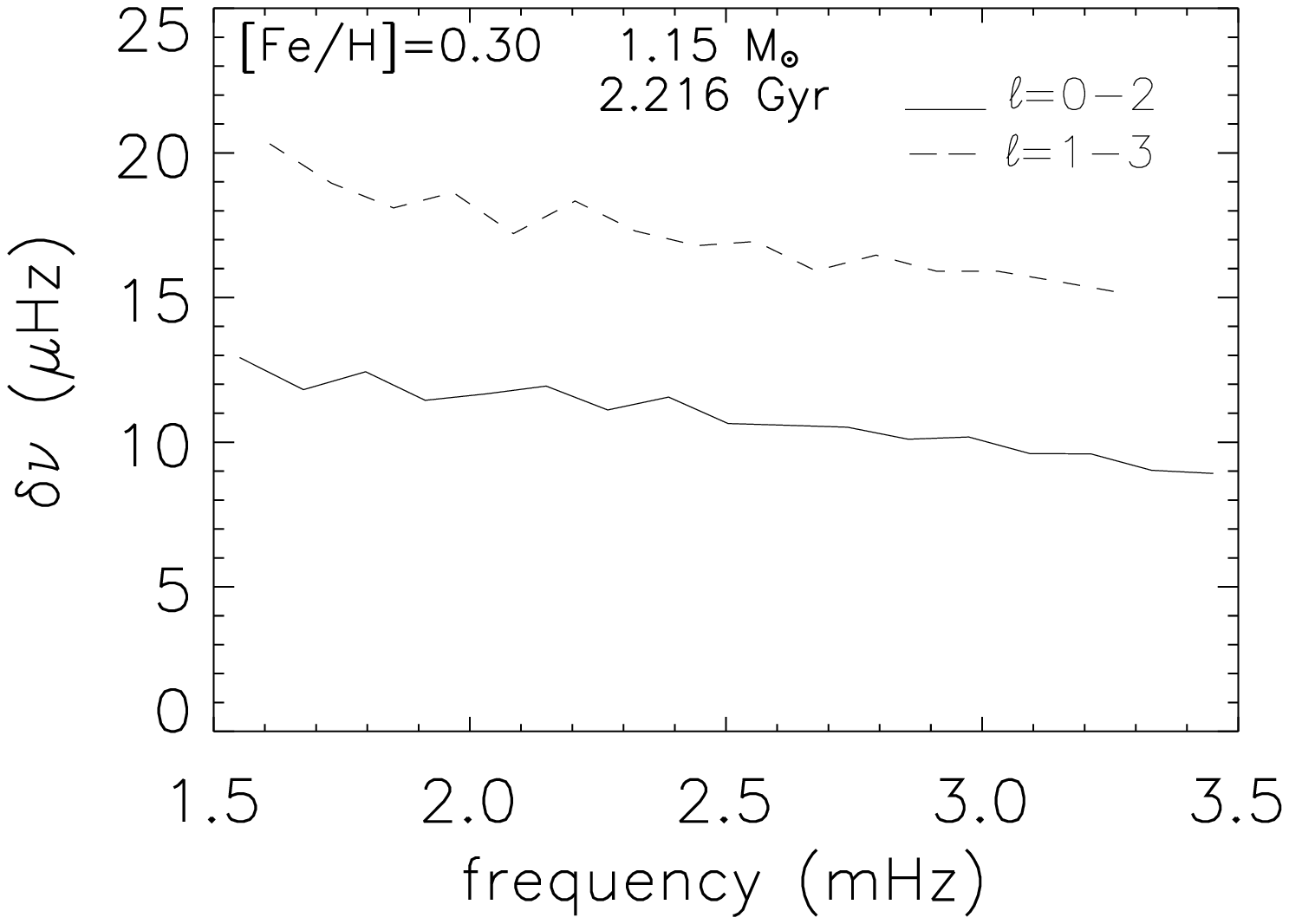}\includegraphics[angle=0,totalheight=5.5cm,width=8cm]{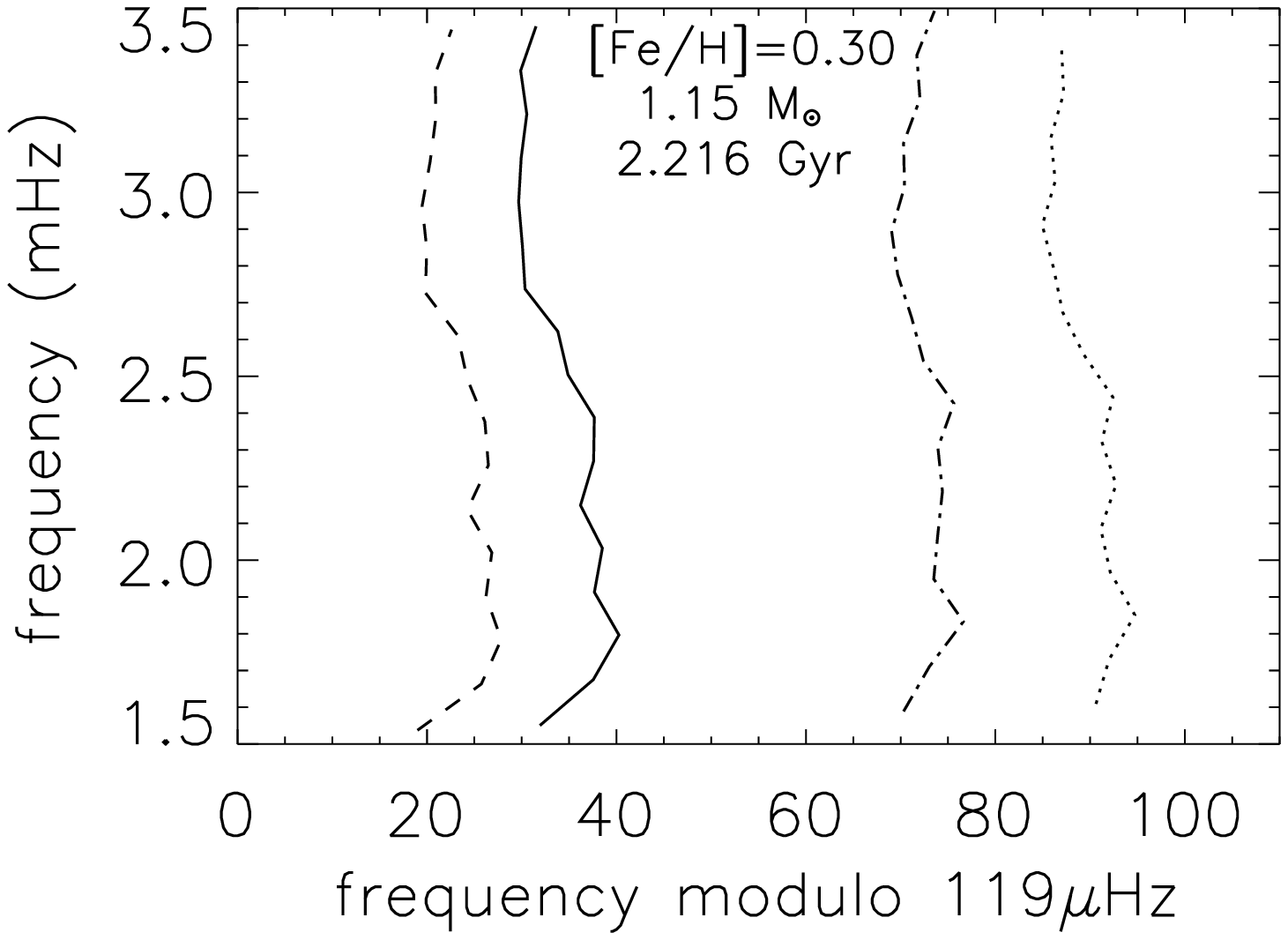}
\includegraphics[angle=0,totalheight=5.5cm,width=8cm]{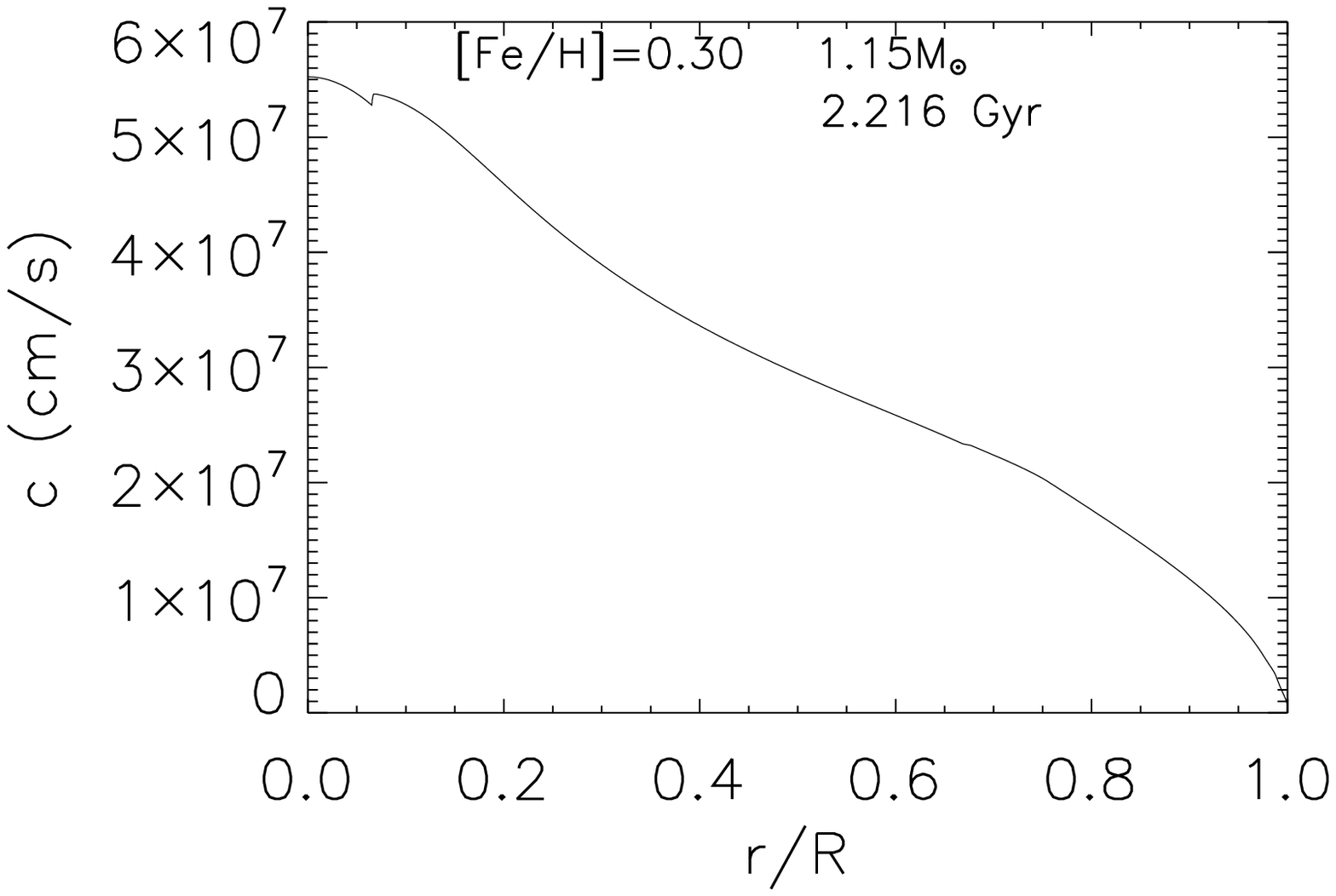}\includegraphics[angle=0,totalheight=5.5cm,width=8cm]{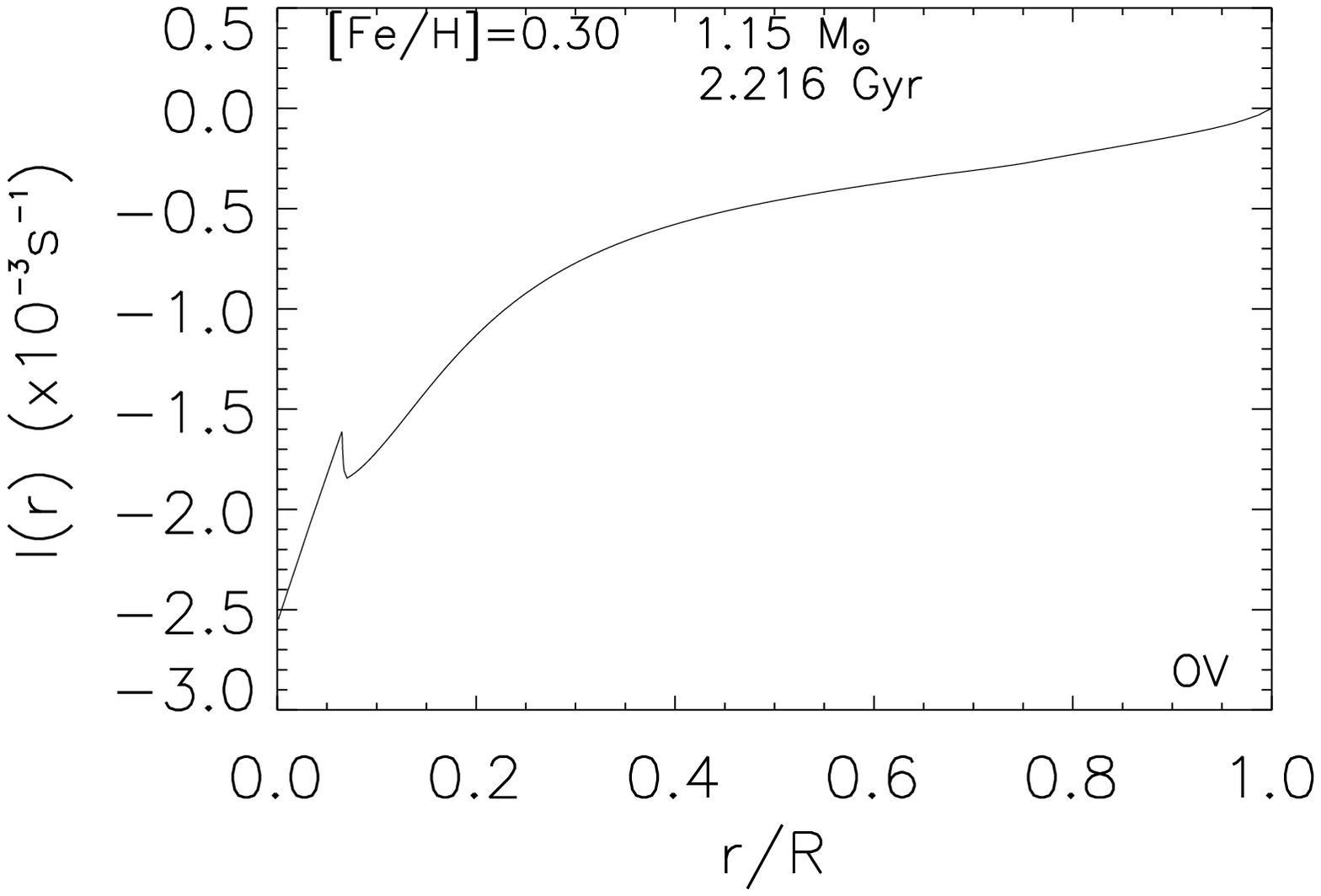}
\end{center}
\caption{Small separations (upper-left panel), echelle diagram (upper-right panel), sound speed profile (lower-left panel), and integral $I(r)=\int_{r_t}^R \frac{1}{r} \frac{dc}{dr} dr$ (lower-right panel) for an overmetallic model ([Fe/H]=0.30) of 1.15~\msol and 2.216~Gyr with overshooting. For the echelle diagram, solid lines are for $\ell=0$, dotted lines for $\ell=1$, dashed lines for $\ell=2$, and dotted-dashed lines for $\ell=3$.}
\label{fig8}
\end{figure*}

\begin{figure*}
\begin{center}
\includegraphics[angle=0,totalheight=5.5cm,width=8cm]{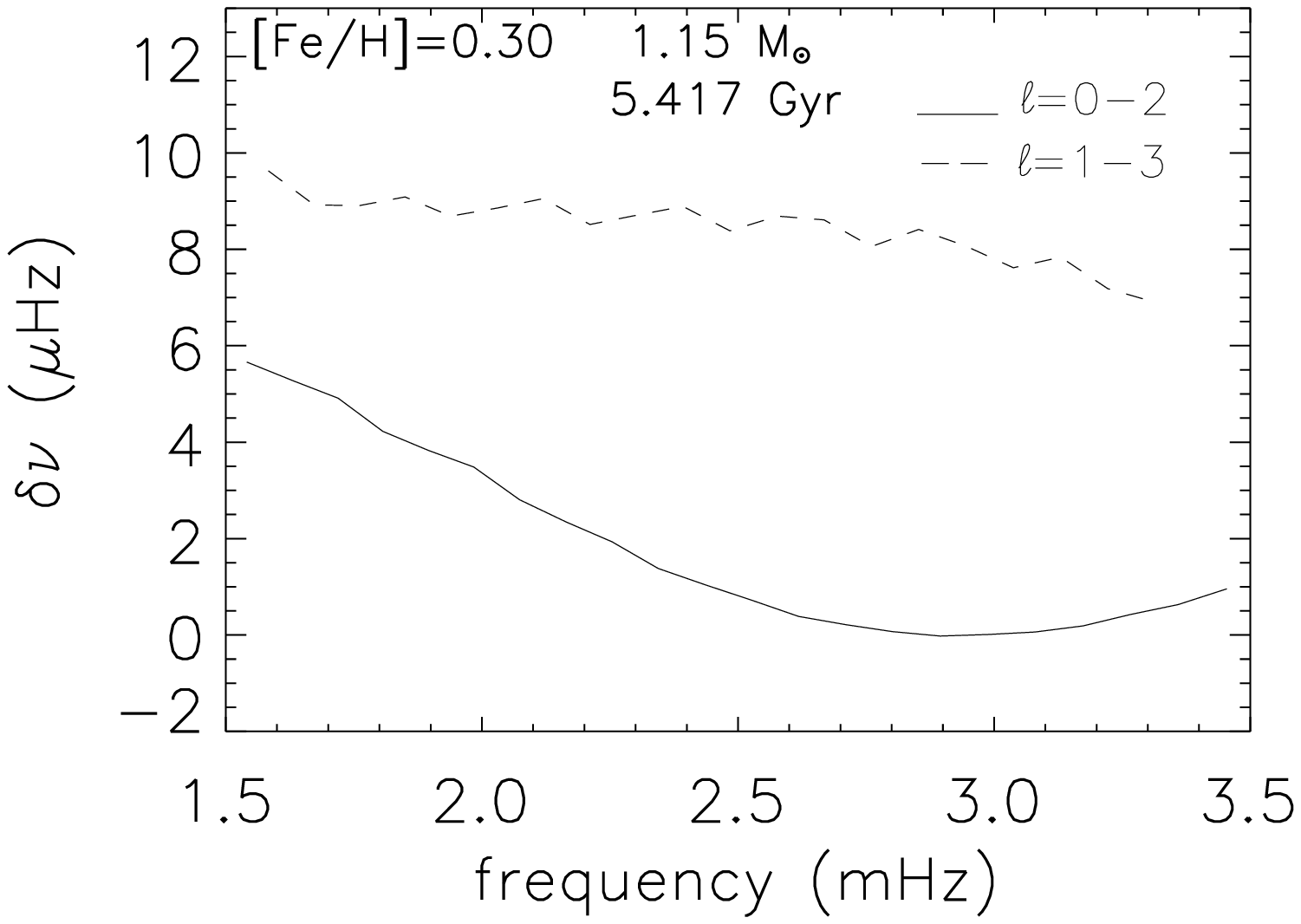}\includegraphics[angle=0,totalheight=5.5cm,width=8cm]{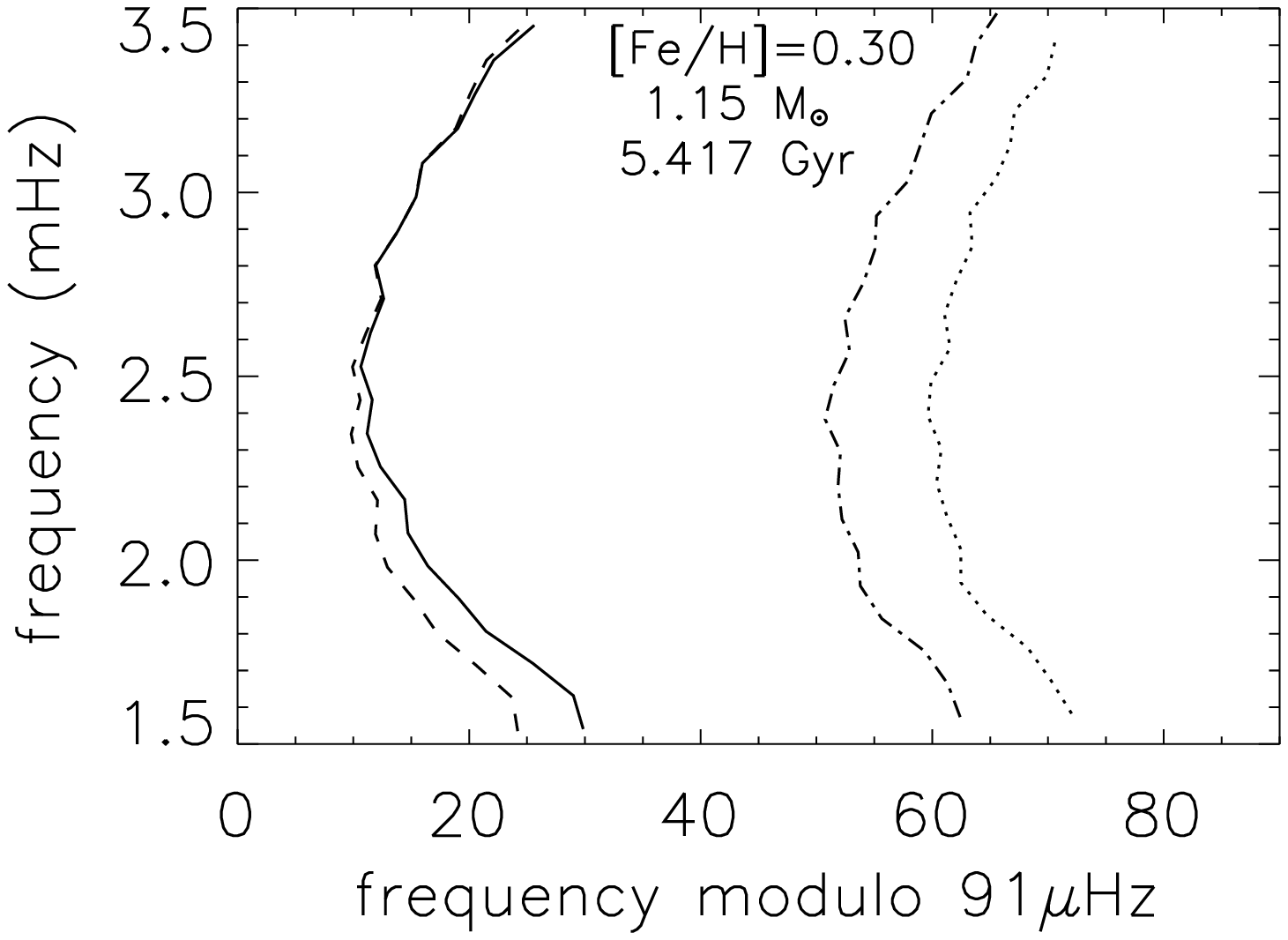}
\includegraphics[angle=0,totalheight=5.5cm,width=8cm]{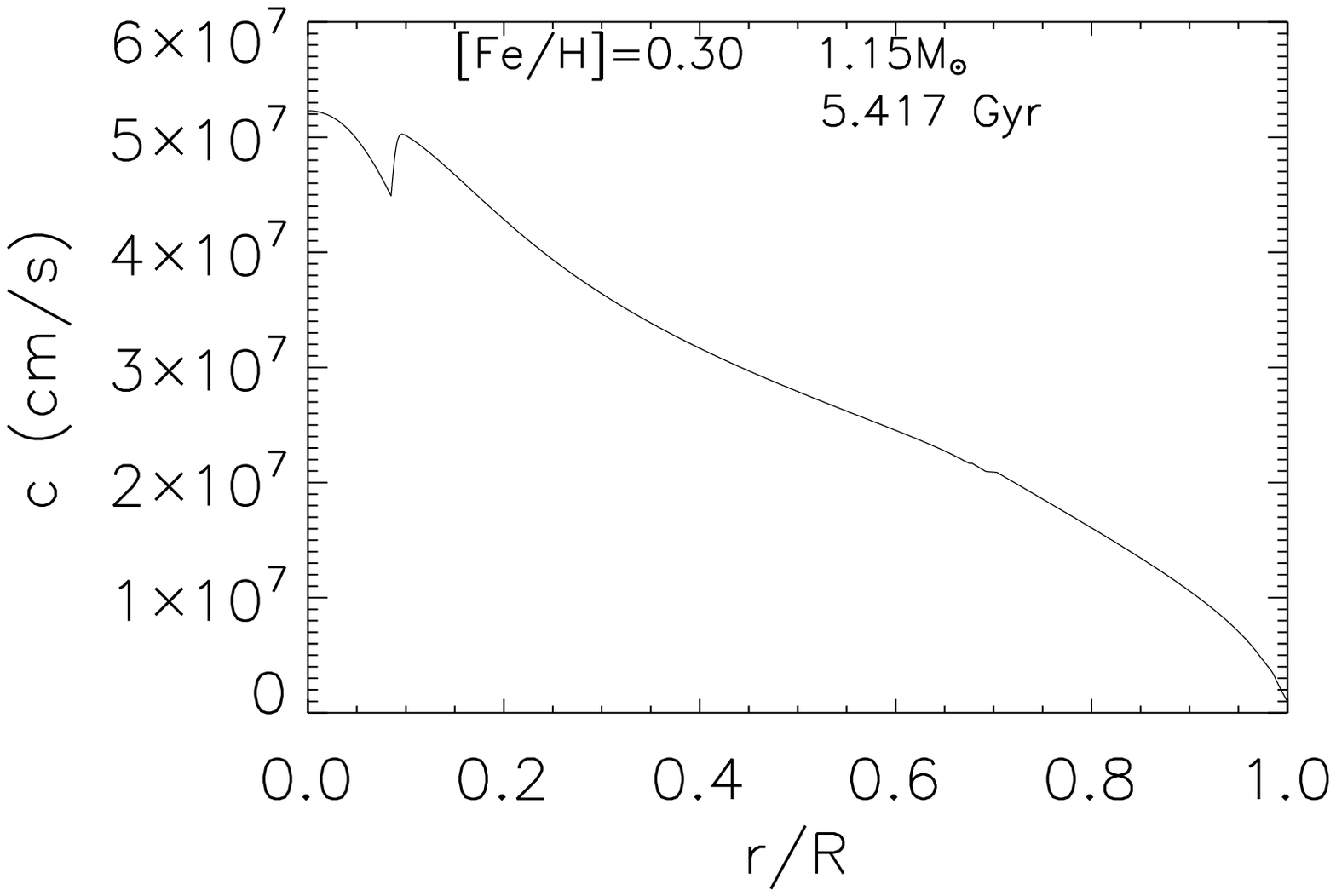}\includegraphics[angle=0,totalheight=5.5cm,width=8cm]{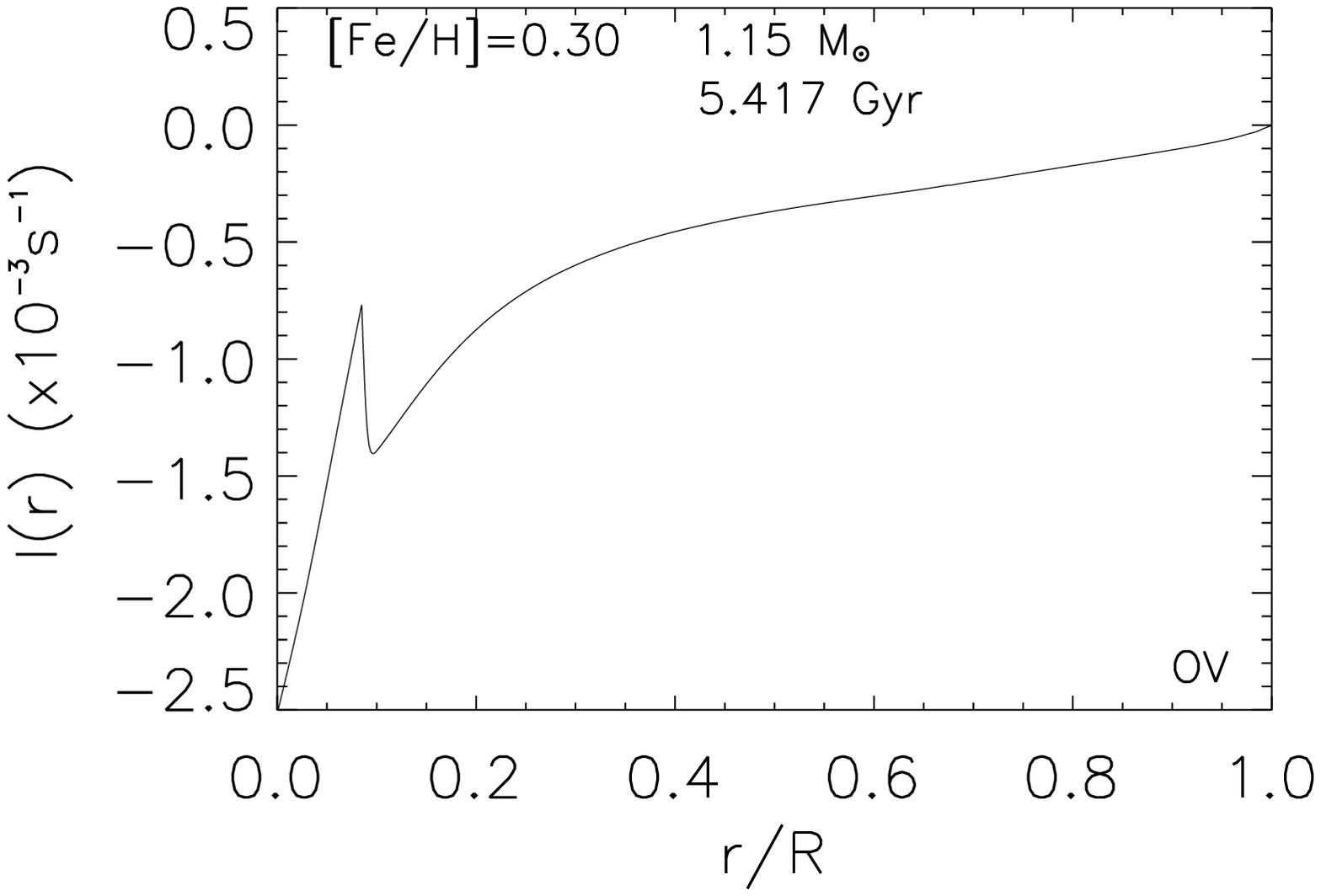}
\end{center}
\caption{Small separations (upper-left panel), echelle diagram (upper-right panel), sound speed profile (lower-left panel), and integral $I(r)=\int_{r_t}^R \frac{1}{r} \frac{dc}{dr} dr$ (lower-right panel) for an overmetallic model ([Fe/H]=0.30) of 1.15 \msol and 5.417 Gyr, with overshooting. It is the first model for which we obtain negative small separations below 3.5 mHz. For the echelle diagram, solid lines are for $\ell=0$, dotted lines for $\ell=1$, dashed lines for $\ell=2$, and dotted-dashed lines for $\ell=3$.}
\label{fig9}
\end{figure*}

\begin{figure*}
\begin{center}
\includegraphics[angle=0,totalheight=5.5cm,width=8cm]{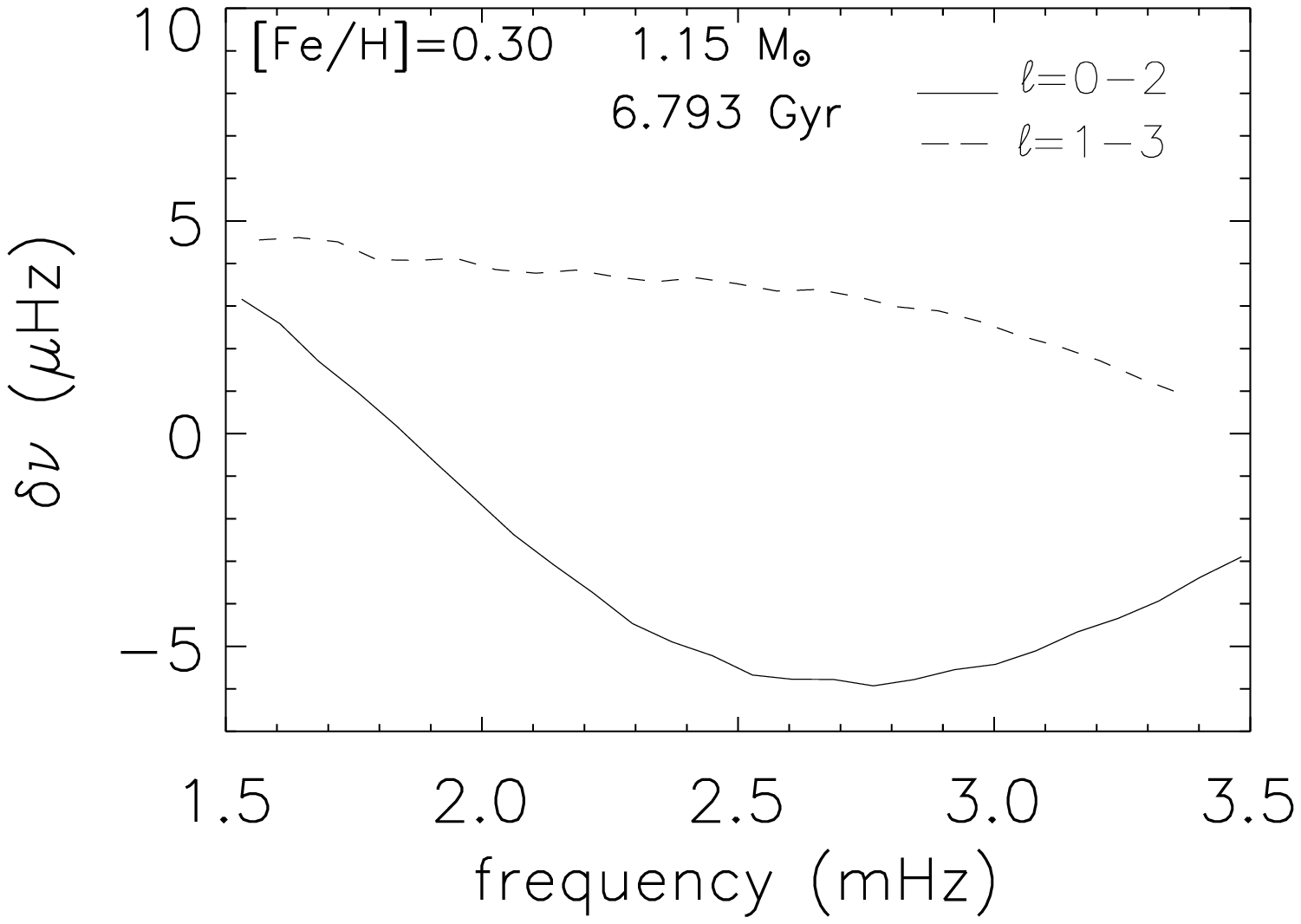}\includegraphics[angle=0,totalheight=5.5cm,width=8cm]{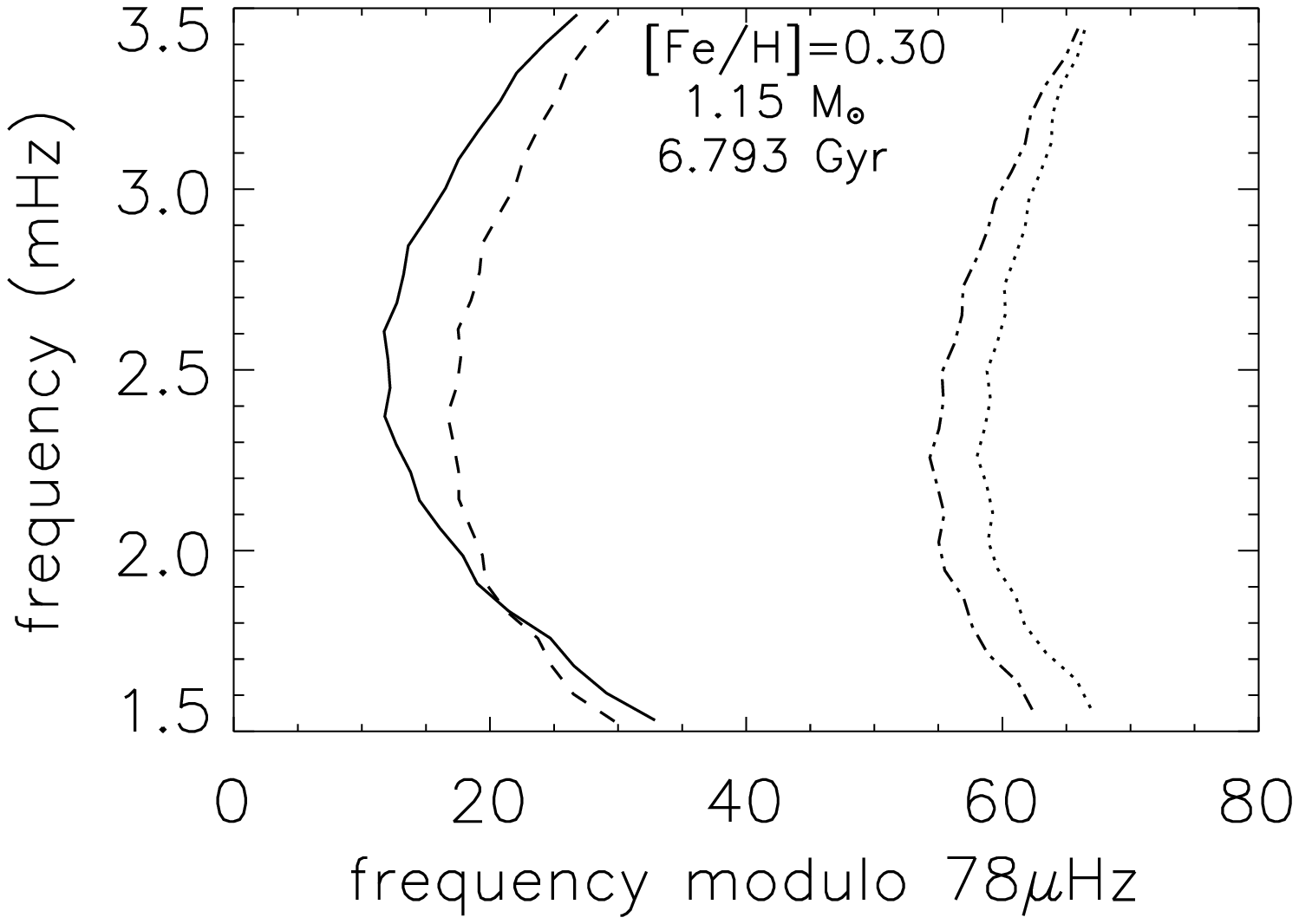}
\includegraphics[angle=0,totalheight=5.5cm,width=8cm]{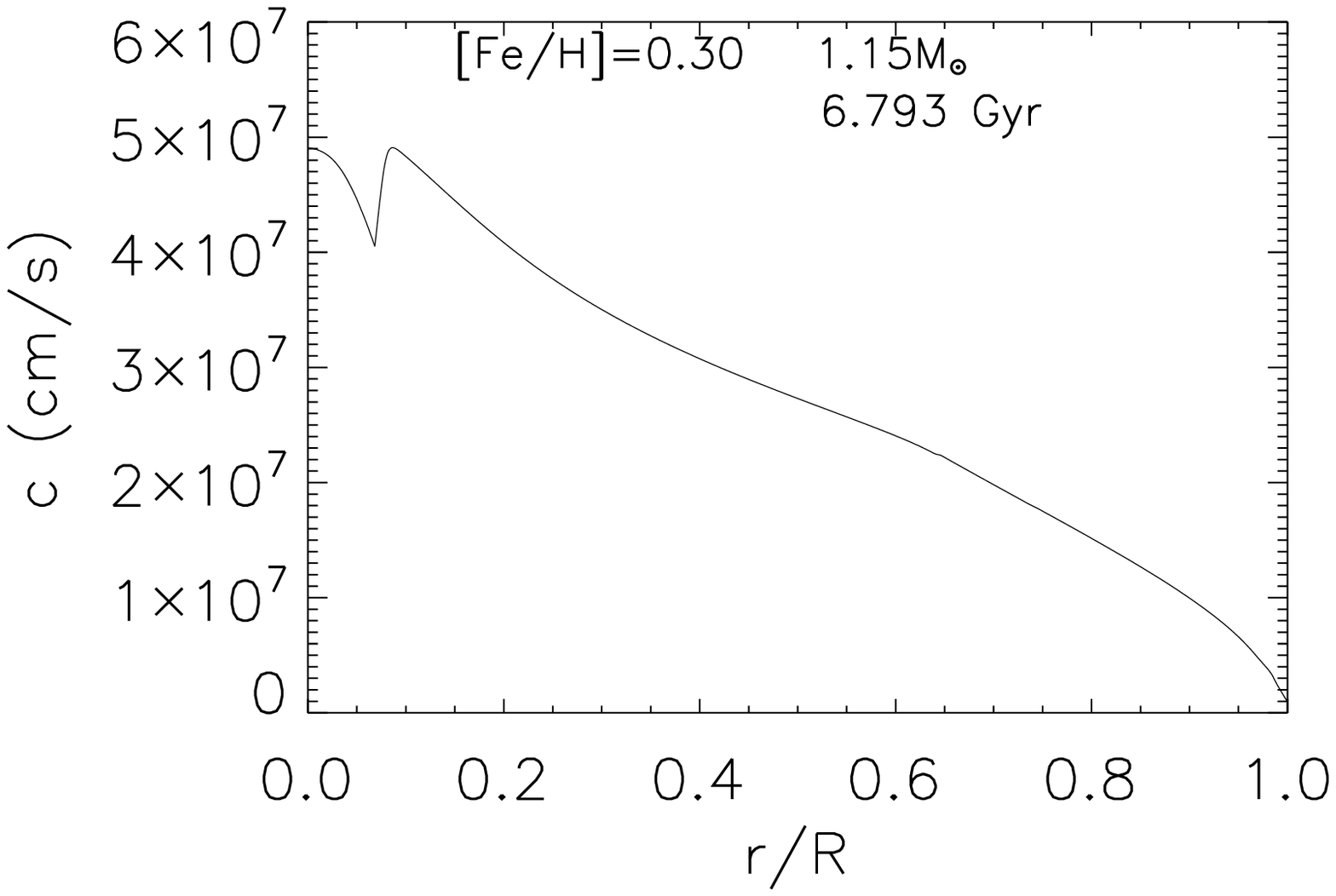}\includegraphics[angle=0,totalheight=5.5cm,width=8cm]{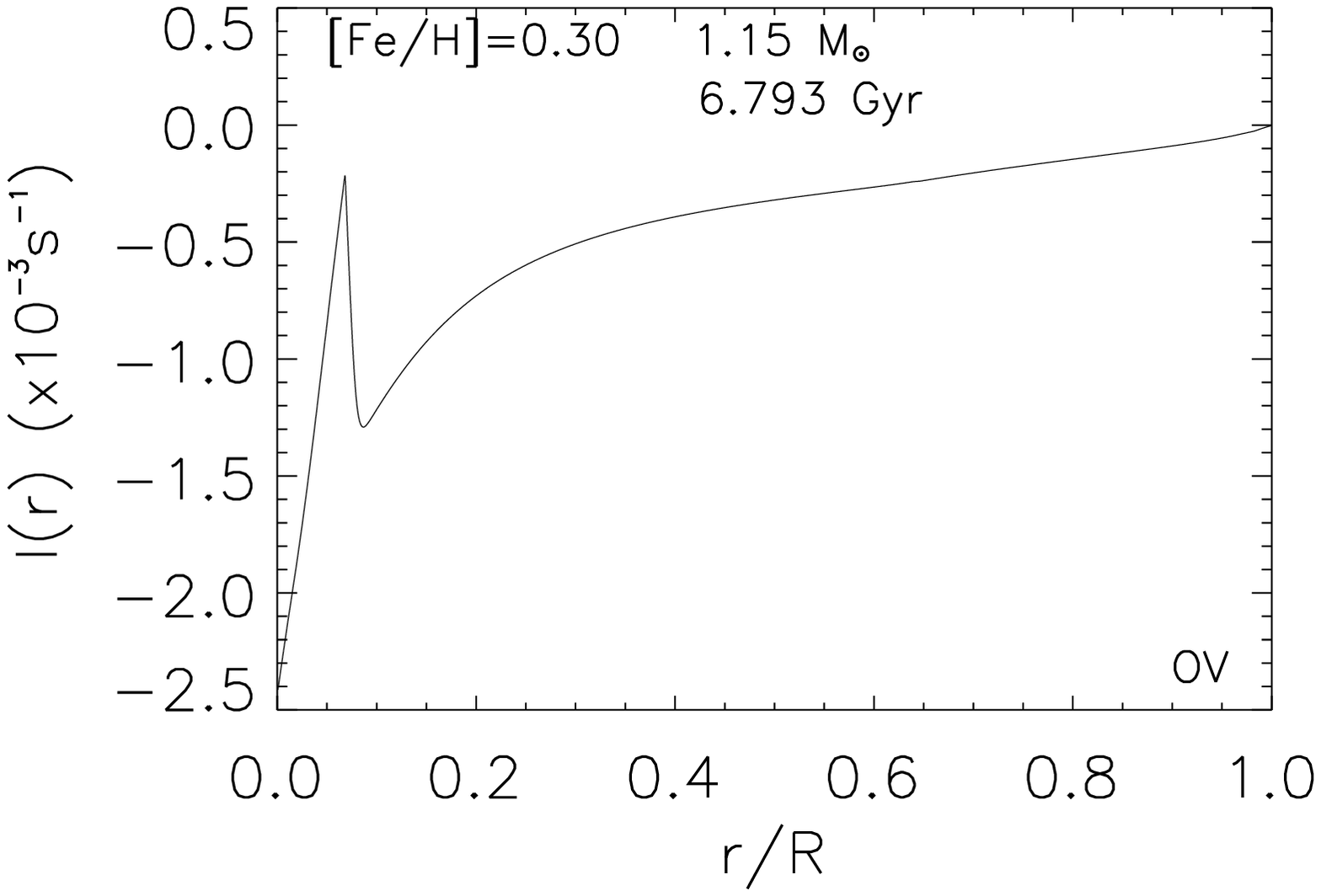}
\end{center}
\caption{Small separations (upper-left panel), echelle diagram (upper-right panel), sound speed profile (lower-left panel), and integral $I(r)=\int_{r_t}^R \frac{1}{r} \frac{dc}{dr} dr$ (lower-right panel) for an overmetallic model ([Fe/H]=0.30) of 1.15 \msol and 6.793 Gyr, without overshooting. This model is on the subgiant branch, with a helium-rich core. For the echelle diagram, solid lines are for $\ell=0$, dotted lines for $\ell=1$, dashed lines for $\ell=2$, and dotted-dashed lines for $\ell=3$.}
\label{fig10}
\end{figure*}

As discussed in Sect.~2, overshooting is added as an extension of the convective core, with a thickness equal to 0.20 times the pressure height scale. 
Figures~\ref{fig8}, \ref{fig9}, and \ref{fig10} present the internal parameters of stars along an evolutionary track of 1.15 \msol, [Fe/H]=0.30 with overshooting.
These figures are directly comparable to Figures~\ref{fig5}, \ref{fig6}, and \ref{fig7}, which presented the same parameters for stars without overshooting. When overshooting is added, the radius of the mixed central core (convection plus overshoot) is increased so that the negative small separations appear for lower frequencies.

The overshooting cases are presented in Tables~\ref{tab4} to \ref{tab6}. The transition models appear at similar ages as for the cases without overshooting. However, as can be seen in Fig.~\ref{fig2}, this corresponds to a relatively earlier stage on the main sequence, as the evolution time scales are longer in the case of larger mixed zones in the stellar centre.

In the overshooting case, due to the larger core radius, the oscillations predicted for the small separations (Bazot \& Vauclair \cite{bazot04} and references therein) begin to appear clearly (Figs.~\ref{fig9} and \ref{fig10}). For example, in the case of 1.15 \msol, 6.793 Gyr (Fig.~\ref{fig10}), the small separations for $\ell=0$~-~$\ell=2$ begin to decrease for increasing frequencies, become negative at $\nu=1.8$ mHz, go on decreasing, go through a minimum at 2.7 mHz, and then begin to increase. They eventually become positive again, for a frequency higher than 3.5 mHz.

This behaviour is related to partial reflexion of the waves inside the mixed core, and the modulation period (in the frequency space) is of the order of the inverse of the acoustic time corresponding to the core boundary (given in the last column of Tables~\ref{tab1} to \ref{tab6}).That we only observe this behaviour for models with overshooting is due to the increase in the core size leading to a decrease in the modulation period.

\section{Summary and discussion}

As powerful as asteroseismology may be, it will always give much less precise results about stellar internal structure than helioseismology does for the Sun. While more than ten million modes can be identified for the Sun, only tens of modes may be observed for stars. In this framework, skilled recipes are needed to better characterise the stellar structures from the observed acoustic frequencies. Among the frequency combinations that have been proposed to test stellar interiors, the ``small separations'', which represent the frequency differences between modes $\ell$, $n$, and $\ell+2$, $n-1$, are known to give important hints about the stellar cores.

In the present paper, we have shown that the small separations, which should, in a first approximation, have a positive sign, may become negative during stellar evolution. Moreover, this behaviour occurs systematically for solar-type stars and may become observable at the end of the main-sequence or at the beginning of the subgiant branch for the $\ell=0$~-~$\ell=2$ separations. It also occurs for the $\ell=1$~-~$\ell=3$ separations, at higher frequencies that are generally outside the observing range. For this reason, we concentrated on the $\ell=0$~-~$\ell=2$ ones. We found that this effect is directly related to the presence of a helium core, itself induced by convection in the central stellar regions. Although convective cores by themselves lead to a significant decrease in the sound velocities in the central stellar regions, the resulting gradients are not sufficient to inverse the sign of the small separations, until enough helium has accumulated. At the end of the main sequence, or at the beginning of the subgiant branch, the chemical gradient induces a much larger sound velocity gradient than convection alone and leads to negative small separations in the observing frequency region. When the star lies at the beginning of the subgiant branch, the convective core has disappeared but the effect continues due to the remaining helium core. From then on, this behaviour goes on, until the stellar structure becomes strongly modified on the subgiant branch. 

The occurrence of negative small separations stems from the modes of different $\ell$ not seeing exactly the same stellar regions. In particular, while the $\ell=0$ modes travel down to the stellar centre, the $\ell=2$ modes travel between the stellar surface and a ``turning point'' that depends on its frequency. Modes with low frequencies may penetrate the stellar core, whereas modes with high frequencies stay outside. The $\ell=2$ modes for which the turning point corresponds to the edge of the stellar core play a special role in this respect. In most cases, negative small separations begin to occur at the frequency of these modes. 

The interest of this behaviour of the small separations is evident, as it may help for characterising convective and helium cores and for constraining overshooting, at least for stars at the end of the main sequence or the beginning of the subgiant branch. However, this may not be very simple in real life. When we observe the acoustic spectra of a solar-type star, the identification of the modes is not easy. That there may be a mode crossing point adds a new difficulty in this respect. Once an observed ``echelle diagram'' has been obtained, it may be difficult to decide whether it shows a crossing point or not.

We will discuss these observational versus theoretical points in a forthcoming paper, in which we will give a new analysis of the star $\mu$ Arae (after Bazot et al. \cite{bazot05}). We will show that models with crossing points for the $\ell=0$~-~$\ell=2$ modes can be obtained if overshooting is added to the convective core. However, only very specific models may fit the overall echelle diagram, so that the constraints are really strong. We will also show that, in the presence of rotational splitting, the modes can be separated without ambiguity, as $\ell=0$ modes cannot show the splitting effects.

As a conclusion, the negative small separations are quite a helpful tool for characterising convective and helium cores, and they may give strong constraints for the possible overshooting at the edge of the convective cores.

\end{document}